%% file: main.tex
\documentclass[runningheads]{llncs}
\usepackage[T1]{fontenc}
\usepackage{graphicx}
\usepackage{algorithm}
\usepackage[noend]{algpseudocode}
\usepackage{booktabs}
\usepackage{multirow}

\newif\iflncs
\lncsfalse   

\input{98_macros}

\newcommand{\Span}{\mathrm{span}}

\begin{document}
\title{End-to-End Hard-Label Cryptanalytic Model Extraction Using Efficient Sign Recovery}
%
%
 \author{Akira Ito\orcidID{0000-0002-4602-7570} \and
 Takayuki Miura\orcidID{0000-0001-8694-312X} \and
 Yosuke Todo\orcidID{0000-0002-6839-4777}}
%
%
  \institute{
 Tohoku University \\ 
   2--1--1 Katahira, Aoba-ku, Sendai-shi, 980-8577, Japan \\
  \email{akira.ito.b1@tohoku.ac.jp}, 
  \and
  NTT Social Informatics Laboratories, \\ 
  3--9--11 Midori-cho,
   Musashino-shi, Tokyo, 180-8585, Japan \\
  \email{tkyk.miura@ntt.com},
  \email{yosuke.todo@ntt.com}
  }
\maketitle              
\pagestyle{plain} 
\begin{abstract}
The importance of deep neural networks (DNNs) is widely recognized, and the parameters obtained through training are regarded as valuable assets.
Recently, attacks that extract these parameters using only oracle queries to a DNN have been actively studied at IACR conferences.
The hard-label setting is the most challenging setting for model extraction, where an adversary can observe only the final output label, such as ``dog'' or ``cat''.
At Eurocrypt 2025, Carlini et al. proposed polynomial-time hard-label extraction of ReLU-based MLPs.
However, one step of this attack process, i.e., sign recovery, requires a large number of queries and substantial computation.
Implementing this step in a black-box setting remains difficult.
Consequently, a fully black-box end-to-end demonstration on trained deep ReLU MLPs has remained a challenge.
In this paper, we propose a new sign-recovery algorithm based on a completely different principle from the existing method.
Our method requires no dedicated queries for sign recovery.
In our experiments, it achieves higher sign-recovery accuracy than the existing method.
Consequently, it enables efficient sign recovery even for trained models.
With our sign-recovery algorithm, all steps of hard-label model extraction can be implemented in a black-box setting.
By combining these implementations, we demonstrate end-to-end model extraction from models trained on MNIST and Fashion-MNIST, with width 16 and 4 or 6 hidden layers, achieving over $98\%$ label agreement.


\end{abstract}
%
%
%

\input{01_introduction}
\input{02_preliminary}

\input{03-sign}
\input{04_e2e}

\section{Conclusion}
We analyzed obstacles to cryptanalytic hard-label model extraction, focusing on sign recovery and also addressing signature recovery, and proposed methods to overcome them.
For sign recovery, we highlighted the implementation challenges and large number of \intersectionspaces required by the boundary-walking method, and proposed cosine and projection-based methods that use simpler computations and fewer \intersectionspaces.
For signature recovery, we addressed spurious groups accepted by consistency checks and unreachable weights in the hard-label setting.
Combining these methods, we demonstrated end-to-end extraction of trained ReLU models with four and six hidden layers of width 16, recovering all neurons except dead and almost-dead neurons and achieving over $98\%$ label agreement on standard Gaussian inputs.

Although our methods limit the additional queries needed for signature and sign recovery, collecting the required \intersectionspaces still incurs a substantial query cost.
Making this collection process more efficient remains an important direction for future work.

%

\subsubsection*{\ackname}
We acknowledge the use of OpenAI Codex (GPT-5.6) to assist with code implementation, debugging, and code review. All research ideas, experimental design, and analyses were developed and verified by the authors.

%
%

\bibliographystyle{splncs04}
\bibliography{main}

\iflncs
\else
\clearpage
\appendix
\input{90_appendix}
\input{91_on_signature}
\input{92_e2e_appendix}
\fi
\end{document}

%% file: 98_macros.tex
\usepackage{xspace}
\usepackage{comment}

\usepackage{subcaption}
\usepackage[whole]{bxcjkjatype}
\usepackage{amsmath}
\usepackage{mathtools}
\usepackage{amssymb,amsfonts}
\usepackage[hypertexnames=false,setpagesize=false]{hyperref}
\usepackage[nameinlink,capitalize,noabbrev]{cleveref}
\usepackage{autonum}
\usepackage{bm}
\usepackage{color}
\usepackage{tabularx}

\crefname{equation}{Eq.}{Eqs.}
\crefname{figure}{Fig.}{Figs.}

\newcommand{\R}{\mathbb{R}}

\newtheorem{model}{Model}

\newcommand{\intersectionpoint}{intersection point\xspace}
\newcommand{\intersectionspace}{intersection space\xspace}
\newcommand{\intersectionpoints}{intersection points\xspace}
\newcommand{\intersectionspaces}{intersection spaces\xspace}

\newcommand{\E}{\mathbb{E}}

\newcommand{\bma}{\bm{a}}
\newcommand{\bmb}{\bm{b}}

\newcommand{\bme}{\bm{e}}

\newcommand{\bmh}{\bm{h}}

\newcommand{\bmn}{\bm{n}}

\newcommand{\bms}{\bm{s}}

\newcommand{\bmu}{\bm{u}}
\newcommand{\bmv}{\bm{v}}
\newcommand{\bmw}{\bm{w}}
\newcommand{\bmx}{\bm{x}}
\newcommand{\bmy}{\bm{y}}

\newcommand{\bfA}{\mathbf{A}}

\newcommand{\bfC}{\mathbf{C}}
\newcommand{\bfD}{\mathbf{D}}

\newcommand{\bfI}{\mathbf{I}}

\newcommand{\bfN}{\mathbf{N}}

\newcommand{\bfP}{\mathbf{P}}

\newcommand{\bfT}{\mathbf{T}}
\newcommand{\bfU}{\mathbf{U}}

\newcommand{\bfW}{\mathbf{W}}

\newcommand{\bfGamma}{\mathbf{\Gamma}}

\renewcommand{\ker}{\mathrm{Ker}}

%% file: 01_introduction.tex
\section{Introduction}
Deep neural networks (DNNs) have brought substantial advances across domains such as image processing, audio signal processing, and natural language processing~\cite{DBLP:conf/nips/VaswaniSPUJGKP17,DBLP:conf/ssw/OordDZSVGKSK16,DBLP:conf/icml/RadfordKHRGASAM21}. 
Nowadays, AI-based services have become an important part of modern digital infrastructure.
However, training these models requires substantial computational resources and large datasets, so the learned parameters are typically treated as confidential intellectual property. If an attacker extracts the parameters of a trained model, it poses a significant risk to the model provider~\cite{DBLP:conf/icml/MartinelliSGB24,DBLP:conf/uss/JagielskiCBKP20,DBLP:conf/nips/FoersterMSH24,DBLP:conf/iclr/DanielyG23,DBLP:conf/fat/MilliSDH19,DBLP:conf/uss/TramerZJRR16,DBLP:conf/icml/RolnickK20,DBLP:conf/sectl/MiuraSY24}. 

Recently, cryptanalytic approaches to model extraction have attracted attention at IACR venues, motivated by structural similarities between multilayer perceptrons (MLPs) and symmetric-key ciphers~\cite{DBLP:conf/crypto/CarliniJM20,DBLP:conf/eurocrypt/CanalesMartinezCHRSS24,DBLP:conf/asiacrypt/ChenDGSWW24,DBLP:conf/eurocrypt/CarliniCHRS25,DBLP:conf/eurocrypt/LiuSELBP26,DBLP:conf/crypto/ItoMT26}. 
Both can be viewed as compositions of linear and nonlinear transformations; the hidden information is a cryptographic key in one case and neural network parameters in the other. 
From this perspective, cryptanalytic techniques that adaptively query an oracle to recover secret information can be naturally applied to model extraction. 

\begin{figure}[t]
    \centering
    \includegraphics[width=\linewidth]{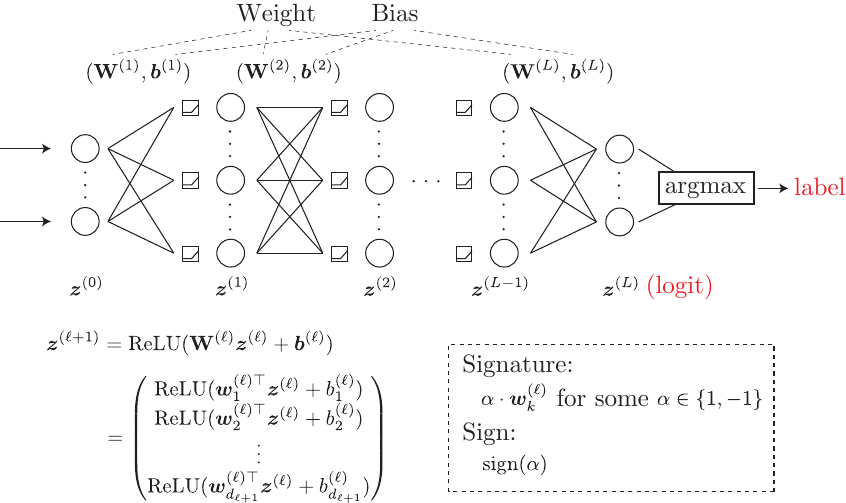}
    \caption{ReLU-based MLP and model extraction. }
    \label{fig:overview}
\end{figure}
This line of research was initiated by Carlini et al.~\cite{DBLP:conf/crypto/CarliniJM20}, who focused on ReLU-based MLPs because of their widespread use. 
\Cref{fig:overview} shows the ReLU-based MLP. 
In a canonical representation of a ReLU-based MLP, the adversary's goal is to extract the \emph{signature}, \emph{sign}, and \emph{bias} of each target neuron layer by layer. 
The original CRYPTO 2020 work assumed that the adversary can observe raw outputs (logits), and this setting is called the \emph{soft-label setting}. 
In the soft-label setting, a polynomial-time algorithm was proposed at Eurocrypt 2024~\cite{DBLP:conf/eurocrypt/CanalesMartinezCHRSS24}.
The feasibility of end-to-end extraction was subsequently demonstrated on trained models at Eurocrypt 2026~\cite{DBLP:conf/eurocrypt/LiuSELBP26}.

The hard-label setting is a different and more restrictive setting.
Logits, which are visible in the soft-label setting, are usually not exposed to users, and only the output label is visible. 
In the hard-label setting, the adversary attempts to recover internal parameters using only these labels. 
Thus, the hard-label setting reveals strictly less information than the soft-label setting. 
The initial work on the hard-label setting appeared at Asiacrypt 2024, where sign recovery required exponential cost~\cite{DBLP:conf/asiacrypt/ChenDGSWW24}.
Later, at Eurocrypt 2025, Carlini et al. proposed algorithms with polynomial running time~\cite{DBLP:conf/eurocrypt/CarliniCHRS25}.
They first collect many \emph{\intersectionspaces}.
Each \intersectionspace is formed by the intersection of an activation boundary, where a neuron switches between active and inactive, and a decision boundary, which separates two output labels. 
Each \intersectionspace is represented by an \intersectionpoint, together with the normal vectors of its two adjacent decision boundaries.
Then, the \intersectionspaces associated with the same neuron are grouped. 
From each group, the target neuron's signature and corresponding bias are recovered up to a common sign. 
Finally, this remaining sign ambiguity is resolved using the method that we call the \emph{boundary-walking method}.

\subsubsection{Difficulty of Sign Recovery in the Hard-Label Setting.}
\Cref{fig:existing} illustrates the boundary-walking method.
Starting from an \intersectionpoint $\bms$ for the target neuron, the method pulls the recovered target signature $\alpha \bmw_t$ and its opposite $- \alpha \bmw_t$ back to input-space directions through the transpose $\bfGamma^\top$ of the local linear map. 
It projects each direction onto the corresponding local decision boundary and walks along the boundary.
When it recognizes a bend in the decision boundary, it determines whether the bend is caused by a neuron-state change in a previous layer, called a \emph{past toggle}, or a subsequent layer, called a \emph{future toggle}. 
If it is a past toggle, it recovers the new decision-boundary normal and resumes walking.
The walk continues until the future toggle is detected.
The total distance from the \intersectionpoint to the future toggle is measured.
The method relies on the heuristic observation that the on-side direction, i.e., $\bmw_t$, tends to affect subsequent layers more strongly and hence reaches a future toggle after a shorter distance.
According to \cite{DBLP:conf/eurocrypt/CarliniCHRS25}, sign recovery aggregates the results from $10^2$--$10^3$ \intersectionpoints.

A black-box implementation of the boundary-walking method is challenging\footnote{Recently, the ePrint version of \cite{DBLP:conf/eurocrypt/CarliniCHRS25} was updated to include a black-box end-to-end hard-label model-extraction demonstration~\cite{DBLP:journals/iacr/CarliniCHRS24}. However, this result should be interpreted carefully because their target is an artificially constructed model whose weights are not learned from data. For details, refer to \Cref{appsec:artificial}.}.
We assume that \intersectionpoints and their two adjacent decision-boundary normals have already been recovered with sufficient precision, since signature recovery requires the same information.
Then, it might be possible to walk along a decision boundary and recognize its first bends.
However, it must distinguish a bend caused by a neuron toggle from a false bend caused by estimation noise. 
It also must avoid crossing multiple toggles at once.
Thus, reliable bend detection requires a substantial number of queries.
After a past toggle, the new decision-boundary normal must be recovered again in the black-box setting.
Then, the original decision boundary may remain close to the point immediately after the toggle, which makes accurate recovery of the new normal more difficult.
This precision cannot be compromised, because otherwise the procedure cannot reliably resume walking along the boundary.
Consequently, the boundary-walking method repeatedly invokes query-intensive and computationally expensive subroutines.
For their trained-model experiments, Carlini et al. \cite{DBLP:conf/eurocrypt/CarliniCHRS25} evaluated this procedure through a proof-of-concept white-box implementation.
Then, decision-boundary normals are computed analytically and toggles are detected from internal states without error.
Even their white-box experiments aggregate results from $10^2$--$10^3$ \intersectionpoints to recover the sign.
In a fully black-box setting, errors in estimating normals and detecting bends can require more samples.

\subsection{Our Contribution}
\begin{figure}[t]
    \centering
    \includegraphics[width=\linewidth]{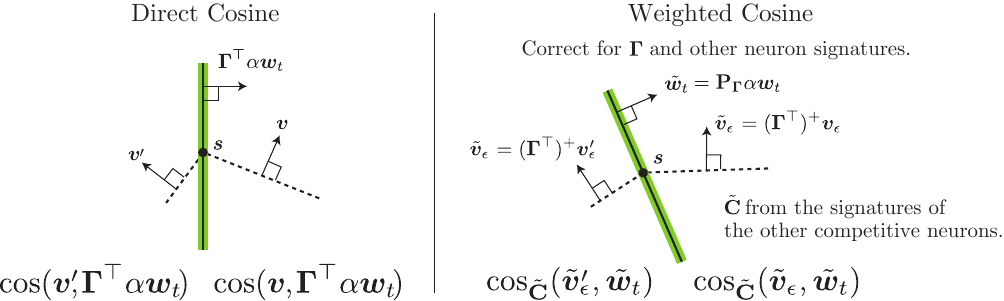}
    \caption{Overview of the new sign-recovery method. }
    \label{fig:comparison}
\end{figure}

\subsubsection{New Sign-Recovery Method: Cosine Method.}

Our main contribution is a new sign-recovery algorithm based on a completely different principle from the boundary-walking method.
Our proposed method, called the \emph{cosine method}, requires neither additional queries nor past/future toggle detection. 
It only performs offline computations on the given \intersectionspaces.
In our experiments, aggregating 50 \intersectionspaces recovers most signs correctly, whereas the boundary-walking method requires substantially more samples. 
Since a certain number of \intersectionspaces have already been collected for signature recovery, in many cases, it requires no additional queries.
It only reuses the collected \intersectionspaces and performs offline computations.

\Cref{fig:comparison} shows the quantity measured by the cosine method.
Concretely, the cosine method compares the cosine similarities between the target signature pulled back to the input space, $\mathbf{\Gamma}^{\top} \alpha \bmw_t$, and the two decision-boundary normals $\bm v$ and $\bm v'$ associated with an \intersectionspace.
We call this simple case the \emph{direct cosine method}.
The sign is recovered from the observation that $\cos(\mathbf{\Gamma}^{\top} \bmw_t, \bmv_{\mathrm{on}})$ has a larger variance than $\cos(\mathbf{\Gamma}^{\top} \bm w_t,\bmv_{\mathrm{off}})$. 
As is clear from this procedure, the proposed method uses no additional queries given \intersectionspaces.

Although comparing cosine similarities is simple, the direct cosine method quickly becomes ineffective in deeper layers.
This is because the local linear map $\mathbf{\Gamma}$ is typically highly distorted, and the cosine similarities become dominated by the anisotropy of $\mathbf{\Gamma}$.
Moreover, in trained models, neuron signatures are anisotropically distributed, and decision-boundary normals themselves can be biased toward particular directions.
We therefore introduce a more advanced \emph{weighted cosine method}.
The weighted cosine method projects the quantities into the image space of the local linear map and corrects them using a matrix determined by the non-target signatures in the same layer. 
Although it requires more computation than the direct cosine method, it still requires no queries, and its computational cost is negligible compared with intersection-point collection. 
Consequently, the proposed method is substantially faster than the existing boundary-walking method, even when compared with its white-box implementation.

How many \intersectionspaces are required to recover a sign using the cosine method?
To answer this question, we estimate the variance of the cosine statistics.
The resulting predictions are experimentally verified on both random models and trained MNIST models.
Our analysis predicts a useful statistical separation at approximately 50 \intersectionspaces under the stated approximations.
On an MNIST-trained model with five hidden layers of width 128, our method correctly recovers 506 of 510 signs in layers 2--5, excluding persistent and dead neurons, using 50 \intersectionspaces per neuron~(\cref{tab:sign-comparison}).
Thus, compared with the boundary-walking method, the proposed method requires no additional queries, has low computational cost, and achieves high accuracy.

\subsubsection{End-to-End Hard-Label Model Extraction.}

Our second contribution is, to the best of our knowledge, the first demonstration of fully black-box end-to-end hard-label model extraction on trained deep ReLU MLPs, enabled by incorporating the sign-recovery algorithm described above.
The feasibility of end-to-end model extraction has already been demonstrated in the soft-label setting~\cite{DBLP:conf/eurocrypt/LiuSELBP26}.
In the hard-label setting, however, end-to-end extraction of trained deep ReLU MLPs has remained a challenge, despite the demonstration on a specially structured artificial model discussed in \Cref{appsec:artificial}.

The obstacles extend beyond sign recovery to signature recovery.
In deeper layers, biased activation patterns can cause \intersectionspaces associated with different neurons to pass the consistency check, producing spurious signatures.
Moreover, the collected \intersectionspaces may not determine every coordinate, leaving some signatures only partially recovered.
Similar issues have been reported in the soft-label setting~\cite{DBLP:conf/eurocrypt/LiuSELBP26}, but addressing them is more difficult when only output labels are observable.
The hard-label setting also introduces a distinct source of false positives: multiple \intersectionspaces can share a decision-boundary hyperplane in the input space of an intermediate layer, causing its normal to be mistaken for a signature.

We analyze these issues and combine methods that address them into an end-to-end attack.
For sign recovery, we use our proposed sign-recovery methods and correct erroneous assignments by checking their consistency with the decision-boundary normals already collected.
For signature recovery, we identify the main cases in which consistency checks produce false positives and reject spurious candidates by examining bends in the decision boundary at candidate activation boundaries and the relationship between candidate signatures and decision-boundary normals.
To recover missing coordinates, we actively search for informative \intersectionspaces and apply existing cross-layer extraction techniques~\cite{DBLP:conf/crypto/ItoMT26} to obtain the missing weight information from activation boundaries in the next layer.

Together, these analyses and methods enable end-to-end extraction from models of width 16 with 4 and 6 hidden layers trained on MNIST and FMNIST.
Our attack pipeline recovers all neurons except dead and almost-dead neurons and achieves over 98\% output agreement with the victim model for every evaluated model on 100,000 inputs drawn from a standard Gaussian distribution.

%% file: 02_preliminary.tex
\section{Preliminaries} \label{sec:preliminaries}

The notation used throughout this paper is summarized in \cref{tab:notation}.

\begin{table}[t]
    \centering
    \caption{Notation}\label{tab:notation}
    \begin{tabularx}{0.95\textwidth}{cX}
        \hline
        Symbol & Description \\ \hline
        $f : \R^{d_0} \to \R^{d_L}$                         & ReLU network with $L$ layers      \\
        $\sigma:\R^{d_*} \to \R^{d_*}$                                            & ReLU function applied component-wise \\
        $\bfW^{(\ell)} \in \R^{d_{\ell} \times d_{\ell-1}}$ & weight matrix in the $\ell$th layer \\
        $\bmb^{(\ell)} \in \R^{d_{\ell}}$                   & bias vector in the $\ell$th layer   \\
        $\bmh^{(\ell)}(\bmx) \in \R^{d_{\ell}}$                   & output vector for input $\bmx$ after the $\ell$th layer \\
        ${\bf\Gamma}_{\bmx}^{(\ell)} \in \R^{d_{\ell} \times d_{0}}$   & forward local linear matrix up to the $\ell$th layer for input $\bmx$    \\
        $\bms \in \R^{d_0}, \quad \mathcal{S} \subset \R^{d_0}$      & \intersectionpoint, \intersectionspace \\
        \hline
    \end{tabularx}
\end{table}

\subsection{ReLU Network and Local Linearity} \label{subsec:relu_network}

\begin{definition}[ReLU~\cite{DBLP:journals/jmlr/GlorotBB11}]
    The ReLU function $\sigma: \mathbb{R}^{d_*} \to \mathbb{R}^{d_*}$ is defined component-wise by
    $\sigma(\bm{x})_i = \max\{\bmx_i, 0\}$ for each $1 \le i \le d_*$. 
    \footnote{Strictly speaking, the definition depends on the input dimension; however, it is standard to use the same symbol $\sigma$ to denote such functions, as they apply the same operation to each component of the input vector.}
\end{definition}

\begin{definition}[ReLU Network]
Let $L\geq2$ be the number of layers, including the affine output layer.
For $1 \le \ell \le L-1$, define 
\begin{align}
\bm{h}^{(\ell)}(\bmx) = \sigma (\mathbf{W}^{(\ell)} \bm{h}^{(\ell-1)}(\bmx) + \bm{b}^{(\ell)}),
\end{align}
where $\bfW^{(\ell)} \in \R^{d_\ell \times d_{\ell-1}}$, $\bmb^{(\ell)} \in \R^{d_\ell}$, and $\bmh^{(0)}(\bmx)=\bmx$.
The input and output of the ReLU function are referred to as the \emph{pre-activation} and \emph{activation}, respectively.
Thus, the pre-activation at layer $\ell$ is $\bfW^{(\ell)}\bmh^{(\ell-1)}(\bmx)+\bmb^{(\ell)}$, and the corresponding activation is $\bmh^{(\ell)}(\bmx)$.
The output layer is defined by
$\bmh^{(L)}(\bmx)=\bfW^{(L)}\bmh^{(L-1)}(\bmx)+\bmb^{(L)}$,
where $\bfW^{(L)} \in \R^{d_L \times d_{L-1}}$ and $\bmb^{(L)} \in \R^{d_L}$.
An $L$-layer \emph{ReLU network} is the function
$f:\R^{d_0}\to\R^{d_L}$ given by
$f(\bmx)=\bmh^{(L)}(\bmx)$.
The parameter set is $\theta = \{(\bfW^{(\ell)}, \bmb^{(\ell)})\}_{\ell=1, \ldots, L}$.
The weight vector of the $k$th neuron in layer $\ell$, denoted by $\bmw_k^{(\ell)}\in\R^{d_{\ell-1}}$, is the transpose of the $k$th row of $\bfW^{(\ell)}$.
For a hard-label model, let $\hat{y}:\R^{d_L}\to[d_L]$ be the function selecting a largest-logit class, with a fixed rule for breaking ties.
The resulting classifier is $\hat{y}\circ f:\R^{d_0}\to[d_L]$.
\end{definition}

A ReLU network partitions the input space into regions according to the activation/inactivation patterns of its neurons, and within each region, the network represents an affine transformation. Such a region is called a \emph{ReLU cell}.
By the positive homogeneity of ReLU, positive neuron-wise rescalings can be propagated through the network without changing the function it represents \cite{DBLP:conf/eurocrypt/CarliniCHRS25}. Hence, in the hard-label setting, we assume without loss of generality that each nonzero row vector of a hidden-layer weight matrix has unit norm, with its bias rescaled by the same factor and the corresponding column of the next-layer matrix adjusted inversely. The final-layer weight matrix is left unnormalized.
We refer to this normalized form as the \emph{canonical representation}.

Since a ReLU network is locally an affine transformation, it exhibits locally linear behavior.
In particular, we define matrices that characterize the local linearity up to each intermediate layer.

\begin{definition}[Forward Local Linear Matrix]
    Let $f : \R^{d_0} \to \R^{d_L}$ be a ReLU network defined as above.
    Let $\bmx \in \R^{d_0}$, and $1 \le \ell \le L$.
    Then, there exist ${\bf\Gamma}_{\bmx}^{(\ell)} \in \R^{d_{\ell} \times d_0}$ and $\bmb_{\bmx}^{(\ell)} \in \R^{d_{\ell}}$ such that
    $\bmh^{(\ell)}(\bmy) = {\bf\Gamma}_{\bmx}^{(\ell)}\bmy + \bmb_{\bmx}^{(\ell)}$
    for any $\bmy$ in the same ReLU cell as $\bmx$.
    The matrix ${\bf\Gamma}_{\bmx}^{(\ell)}$ is called the \emph{forward local linear matrix} at $\bmx$.
    Set ${\bf\Gamma}_{\bmx}^{(0)}=\bfI_{d_0}$ for all $\bmx\in\R^{d_0}$, where $\bfI_{d_0}$ denotes the $d_0\times d_0$ identity matrix.
\end{definition}

\subsection{Cryptanalytic Model Extraction}

The goal of cryptanalytic model extraction is to recover the parameters of a target model as precisely as possible from oracle access to its predictions, up to transformations that preserve its observable behavior.

\paragraph{Adversarial Capabilities and Objectives.}
In line with \cite{DBLP:conf/eurocrypt/CarliniCHRS25}, we consider the following attacker capabilities and goals.
The attacker queries an oracle $\mathcal{O}(\bmx)=\hat y(f_\theta(\bmx))$ on adaptively selected inputs and observes only the corresponding class labels.
The goal is to reconstruct parameters $\hat{\theta}$ that reproduce the target classifier as accurately as possible. Hard-label access does not uniquely determine logits: adding a common affine function to all logits or multiplying them by a common positive scalar preserves every label.
In particular, functional model extraction is defined in terms of functional equivalence.
\begin{definition}[Hard-label functional equivalence]
Let $f_{\theta}$ and $\hat f_{\hat\theta}$ be the target and extracted logit models, and let $\mathcal{D}_X$ be a probability distribution supported on $X\subseteq\R^{d_0}$.
The corresponding classifiers are $\epsilon$-functionally equivalent with respect to $\mathcal{D}_X$ if $\Pr_{\bmx\sim\mathcal{D}_X}
    [\hat y(f_\theta(\bmx))=\hat y(\hat f_{\hat\theta}(\bmx))]
    \geq 1-\epsilon$.
\end{definition}
Accordingly, we evaluate functional extraction by label agreement under a specified input distribution, separately from parameter-recovery accuracy.

Furthermore, consistent with prior work, we adopt the following standard assumptions.
\textbf{Architecture knowledge:} The attacker is aware of the model architecture, including the number of layers and units.
\textbf{Unrestricted inputs:} The attacker can query arbitrary inputs from $\mathbb{R}^{d_0}$.
\textbf{Precise computation:} The DNN is executed with sufficiently high-precision floating-point arithmetic.
\textbf{Fully connected ReLU network:} The target is a multilayer perceptron with fully connected layers, ReLU hidden activations, and an affine output layer.

\subsection{Overview of Hard-Label Model Extraction} \label{subsec:overview_hard_label_model_extraction}

The hard-label model extraction attack proposed by Carlini et al.~\cite{DBLP:conf/eurocrypt/CarliniCHRS25} explores the decision boundary and leverages \emph{\intersectionspaces}, where the decision boundary intersects activation boundaries---i.e., boundaries at which neuron activations switch---to reconstruct neuron parameters.

\begin{definition}[Intersection space and \intersectionpoint]
    For a ReLU network, the intersection between an activation boundary and a decision boundary is termed an \emph{\intersectionspace}. 
    A point lying within an intersection space is referred to as an \emph{\intersectionpoint}.
\end{definition}

Let $\mathcal{S}\subset\R^{d_0}$ be this affine space. It extends the local intersection beyond the adjacent ReLU cells; its distant points need not lie on the network's actual boundaries.
We use two equivalent representations: $(\bms,\bmv,\bmv')$, where $\bms$ is an \intersectionpoint and $\bmv,\bmv'$ are the two adjacent normals, and $(\bms,\bfN)$, where the columns of $\bfN\in\R^{d_0\times(d_0-2)}$ form an orthonormal basis of $\operatorname{span}\{\bmv,\bmv'\}^{\perp}$.

Details on how to collect \intersectionspaces are provided in \cref{appsec:collect_dual_spaces}.

\paragraph{Overview of hard-label model extraction.}
An overview of the attack is as follows.
\begin{enumerate}
\item Collect a large number of \intersectionspaces.
\item Starting from the shallowest layer, repeat the following steps:
\begin{itemize}
\item Identify pairs of \intersectionspaces that correspond to the same activation boundary (\cref{subsec:recovering_the_signature}).
\item Group \intersectionspaces corresponding to the same activation boundary and recover the signature of the neuron associated with that boundary (\cref{subsec:recovering_the_signature}).
\item Recover the signs of the collected signatures (\cref{subsec:sign-recovery}).
\item For signatures that cannot be fully recovered, defer their identification to the cross-layer extraction step \cite{DBLP:conf/crypto/ItoMT26}.
\end{itemize}
\item Proceed layer by layer from shallow to deep layers.
\end{enumerate}

\subsection{Signature Recovery}\label{subsec:recovering_the_signature}

\begin{figure}[t]
\centering
\includegraphics[width=0.90\linewidth]{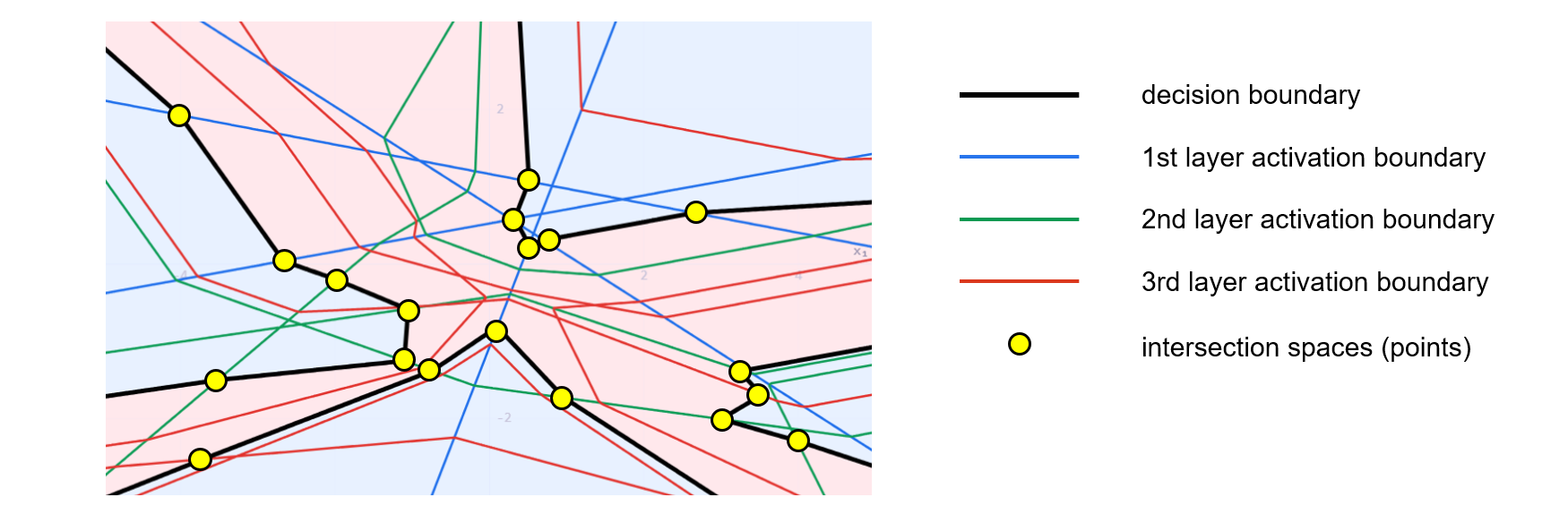}
\caption{Illustration of a two-dimensional input space.}
\label{fig:sameactiveboundary}
\end{figure}

In the input space of its layer, a neuron's weight vector is normal to its activation-boundary hyperplane.
Thus, given multiple \intersectionspaces on the same activation boundary, as illustrated in \cref{fig:sameactiveboundary}, we can reconstruct the boundary and recover a nonzero scalar multiple of the weight vector, which we call its \emph{signature}.

\begin{definition}[Signature]\label{def:signature}
Let $\bmw_k^{(\ell)}\in\R^{d_{\ell-1}}$ be the weight vector of the $k$th neuron in layer $\ell$, i.e., the transpose of the $k$th row of $\bfW^{(\ell)}$.
Its signature is $\alpha\bmw_k^{(\ell)}$ for  $\alpha\in \{1, -1 \}$.\footnote{As discussed earlier, the norm can be normalized without loss of generality, and hence it suffices to consider the values $1$ and $-1$.}
\end{definition}

Signature recovery consists of two stages.
First, a consistency check tests whether two mapped \intersectionspaces satisfy a common affine constraint; the procedure is given in \textsc{IsConsistent}() (\cref{alg:pre-Consistent}) in \cref{appsec:signature-recovery}.
Then, after collecting sufficient consistent \intersectionspaces, the corresponding signature is reconstructed using \textsc{RecoverSignature}() (\cref{alg:pre-signaturerec}) in \cref{appsec:signature-recovery}.


The consistency check is performed as follows.
Let $\mathcal{S}_1$ and $\mathcal{S}_2$ be \intersectionspaces with basis matrices $\bfN_1,\bfN_2$ and representative points $\bms_1,\bms_2$.
For $i\in\{1,2\}$, let $\bfT_i=\bfGamma_{\bms_i}^{(\ell-1)}\bfN_i$ and $\bmu_i=\bmh^{(\ell-1)}(\bms_i)$ be their mapped directions and representative points in the input space of layer $\ell$.
Let $\nu$ be the number of coordinates corresponding to neurons in layer $\ell-1$ that are active at either point; for $\ell=1$, set $\nu=d_0$.
The test accepts if $
    \operatorname{rank}[\bfT_1,\bfT_2,\bmu_1-\bmu_2]<\nu$.
    \label{eq:consistency-test}
This detects a common affine constraint on the active coordinates. It is necessary for a shared activation hyperplane whose normal has a nonzero restriction to these coordinates, but it is not sufficient to identify a common target neuron.
In particular, the spurious groups analyzed in \cref{sec:e2e} can also pass this test.

Now suppose that $\mathcal{S}_1,\ldots,\mathcal{S}_m$ correspond to the same target neuron.
With $\bfT_i$ and $\bmu_i$ defined as above, form
\begin{align}
    \bfU=[\bfT_1,\ldots,\bfT_m,
        \bmu_2-\bmu_1,\ldots,\bmu_m-\bmu_1].
    \label{eq:signature-matrix}
\end{align}
If $\operatorname{rank}(\bfU)=d_{\ell-1}-1$, the one-dimensional null space of $\bfU^\top$ determines the signature up to sign.
Let $\bmw$ be a unit vector in this null space. The corresponding bias is $b=-\bmw^\top\bmu_1$.
The representative-point differences are needed to enforce a common affine hyperplane, in addition to orthogonality to the mapped directions.
If the rank is smaller, the signature is not uniquely determined; additional informative spaces or the methods in \cref{sec:e2e} are needed.
A larger rank indicates inconsistent data in exact arithmetic.
All ranks and null spaces are computed using singular value decomposition (SVD).



\subsection{Sign Recovery}\label{subsec:sign-recovery}

Signature recovery determines a signature $\alpha \cdot \bmw_k^{(\ell)}$ for the $k$th neuron in the $\ell$th layer, and the remaining unknown variable is the sign, $\alpha \in \{1, -1\}$. 
Recovering this sign is necessary to determine which side of the corresponding activation boundary is the active side of the ReLU.


The existing sign-recovery method is based on a heuristic observation: when a carefully chosen direction changes the target neuron's pre-activation by $\epsilon$ from a point on an activation boundary, this perturbation $\epsilon$ affects the subsequent layer if the direction is on the active (on) side, whereas it is zeroed out by ReLU if it is on the inactive (off) side.  
In the soft-label setting \cite{DBLP:conf/eurocrypt/CanalesMartinezCHRSS24}, the sign is successfully recovered by measuring the magnitude of the difference between the output logits. 
Although the same concept is applied in the hard-label setting \cite{DBLP:conf/eurocrypt/CarliniCHRS25}, it measures a more indirect distance because logits are unobservable. 

\begin{figure}[t]
    \centering
    \includegraphics[width=.9\linewidth]{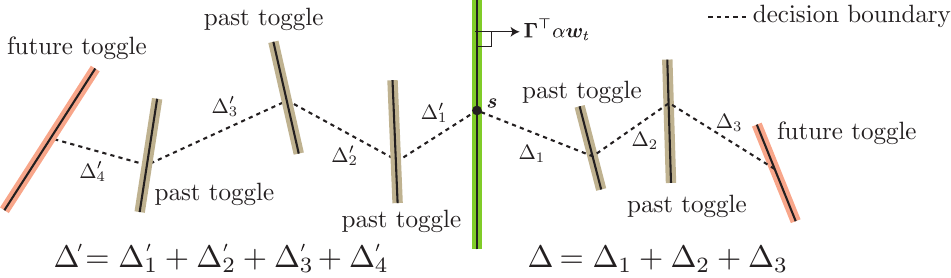}
    \caption{Existing sign-recovery method. Compare the distance between $\Delta$ and $\Delta'$. }
    \label{fig:existing}
\end{figure}

We call the existing method the \emph{boundary-walking method}, and \Cref{fig:existing} summarizes the behavior. 
\begin{enumerate}
    \item Starting from an \intersectionpoint $\bms$, pull the target signature back to the input space\footnote{An alternative is \emph{perfect control}. A direction $\Delta \bmx$ satisfying $\bfGamma \Delta \bmx = \alpha \bme_t$ changes only the target pre-activation. Such a direction exists when $\bme_t \in \mathrm{Im}(\bfGamma)$; full row rank is sufficient. However, the condition is rarely satisfied in deeper layers. } as $(\bfGamma_{\bms}^{(\ell - 1)})^\top \alpha \bmw^{(\ell)}_t$. Project this direction onto the corresponding decision boundary. 
    \item Walk along the decision boundary until a bend is detected. If it is caused by a past toggle, i.e., a neuron state switches before the target layer, recover the new decision-boundary normal after the bend and resume walking.
    \item If a bend in the decision boundary is identified and its cause cannot be explained by past toggles, it supposes future toggles, i.e., a neuron state switches after the target layer. The total distance of observing the future toggle is measured.
    \item Repeat in the opposite direction and compare the two distances. 
\end{enumerate}
In other words, the distance to a future toggle is used instead of the magnitude of the logit difference. 
Considering the underlying heuristic observation, it is expected that the on side will toggle a future neuron's state sooner.

Note that a faster future toggle on the on side does not necessarily hold in all \intersectionspaces, since these movements also change other neurons' outputs in the target layer and they also affect subsequent layers. 
The procedure is therefore repeated over multiple \intersectionpoints corresponding to the same target neuron. 
Two distances are compared for every \intersectionpoint, and a direction with a shorter distance gets one vote. 
If repeated past toggles prevent a reliable comparison, the corresponding \intersectionpoint is discarded.
The authors of \cite{DBLP:conf/eurocrypt/CarliniCHRS25} estimated the number of \intersectionpoints required for high-confidence sign recovery through (white-box) experiments. While 100 points are sufficient for the first layer, approximately 1,000 points are required for the second and subsequent layers.

%% file: 03-sign.tex
\section{Efficient Sign Recovery without Dedicated Queries} \label{sec:sign}

\subsection{Motivation}
Hard-label model extraction by Carlini et al.~\cite{DBLP:conf/eurocrypt/CarliniCHRS25} is conceptually implementable, but its black-box implementation remains challenging. 
In particular, the implementation of the boundary-walking method causes significant difficulty.


First, a decision-boundary normal recoverable in a black-box setting inevitably contains some noise.
The boundary-walking method is highly sensitive to the noise of decision-boundary normals.
When the normal is noisy, even moving along a decision boundary is not straightforward.
One must distinguish a bend caused by a neuron toggle from a false bend caused by noise.
The step size during movement is also important.
An adversary does not know how far it must move before a toggle occurs. Large steps introduce a risk that some neuron toggles are overlooked. Small steps cause the number of queries required before a toggle to explode.
We need to recover the new decision-boundary normal once we recognize a past toggle.
A point immediately after a toggle is also close to the original decision boundary, so additional care is needed to obtain an accurate decision-boundary normal.
Even after all these issues can be resolved efficiently and accurately, the same procedure must be performed on both the on and off sides until the future toggle and be repeated $10^2$--$10^3$ times for each target neuron.
Note that the number of iterations is estimated in a noise-free white-box implementation. 
We might need more points when noise is present.

In practice, the authors of~\cite{DBLP:conf/eurocrypt/CarliniCHRS25} themselves provide only a white-box proof of concept for the trained model. 
The recent end-to-end implementation \cite{DBLP:journals/iacr/CarliniCHRS24} is demonstrated on a specially constructed artificial model that facilitates sign recovery. These results leave the practical cost and robustness of fully black-box boundary walking on trained deep networks unresolved.


\subsection{Intuition Behind the Direct Cosine Method}
\begin{figure}[t]
    \centering
    \includegraphics[width=.9\linewidth]{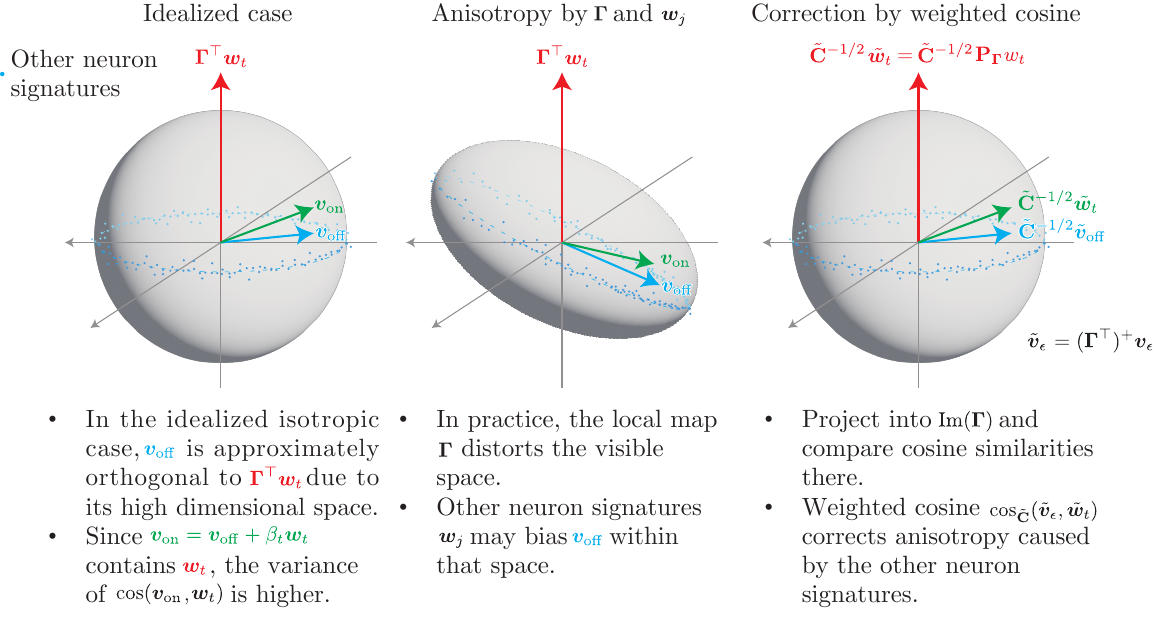}
    \caption{Schematic three-dimensional visualization of the (weighted) cosine methods, obtained by taking the target-signature direction as the $z$-axis.}
    \label{fig:geometric-view}
\end{figure}
Our new sign-recovery algorithm is based on a completely different principle from existing methods. 
Unlike the existing method, it does not require dedicated queries. 
The method requires only \intersectionspaces, which are already known to an adversary for signature recovery. 
Once they are given, just offline computation is sufficient. 

The basic principle is as follows:
We compute the cosine similarities between each decision-boundary normal and the recovered signature.
Our method exploits the heuristic that the cosine similarity with the on-side decision-boundary normal has larger variance than that with the off-side normal. 

Why does the on-side cosine similarity have a higher variance?
Consider recovering the sign of the $t$th neuron in hidden layer $\ell$. 
Let $\bmw_t \in \R^{d_{\ell-1}}$ be the target weight vector, and let $\bmw_j \in \R^{d_{\ell-1}}$ be the $j$th weight vector in the same layer.
At an \intersectionpoint $\bms$, write $\bfGamma=\bfGamma^{(\ell-1)}_{\bms}\in\R^{d_{\ell-1} \times d_0}$. 
Then, using the index set $\mathcal{J}_{\mathrm{act}}$ of neurons active in the same layer, the decision-boundary normal on the off side can be written as
\[
\bmv_{\mathrm{off}} := \bfGamma^\top\sum_{j \in \mathcal{J}_{\mathrm{act}}} \beta_j \bmw_j \in \R^{d_0}.
\]
On the on side, we consider the same decision boundary with only the target neuron state switched, and thus its normal is
\[
\bmv_{\mathrm{on}} := \bmv_{\mathrm{off}} + \beta_t\bfGamma^\top\bmw_t.
\]
Thus, $\bmv_{\mathrm{on}}$ contains the $\bfGamma^\top\bmw_t$ component. 

A direct cosine is defined by 
$
\cos(\bfGamma^\top \alpha_t \bmw_t,\bmv_\epsilon),
$
where $\epsilon \in \{\mathrm{off}, \mathrm{on} \}$. 
Both the target signature $\bfGamma^\top \alpha_t \bmw_t$ and the decision normal $\bmv_{\mathrm{\epsilon}}$ lie in $\operatorname{Im}(\bfGamma^\top)\subseteq\R^{d_0}$. 
The dimension of this subspace is $r_{\bfGamma} := \dim\operatorname{Im}(\bfGamma^\top)=\operatorname{rank}(\bfGamma)$. 

Assuming that $\bfGamma^\top \alpha_t \bmw_t$ and $\bmv_\mathrm{off}$ are independent and isotropic in this subspace, as a typical behavior in high-dimensional space, we have 
\[
\cos(\bfGamma^\top \alpha_t \bmw_t,\bmv_{\mathrm{off}})
\overset{\mathrm{approx}}{\sim}
\mathcal{N}\!\left(0,\frac{1}{r_{\bfGamma}}\right).
\]
The left figure of \Cref{fig:geometric-view} illustrates this idealized geometry. Under the isotropic model, the target direction is nearly orthogonal to the active non-target directions whose linear combination forms the off-side normal. 

The on-side normal additionally contains the target component $\beta_t\bfGamma^\top\bmw_t$. 
Thus, its cosine is expected to have larger variance than that of the off-side.

\subsubsection{Experiments.}
\begin{figure}[t]
  \centering
  \begin{subfigure}[b]{0.48\textwidth}
    \centering
    \includegraphics[width=\textwidth]{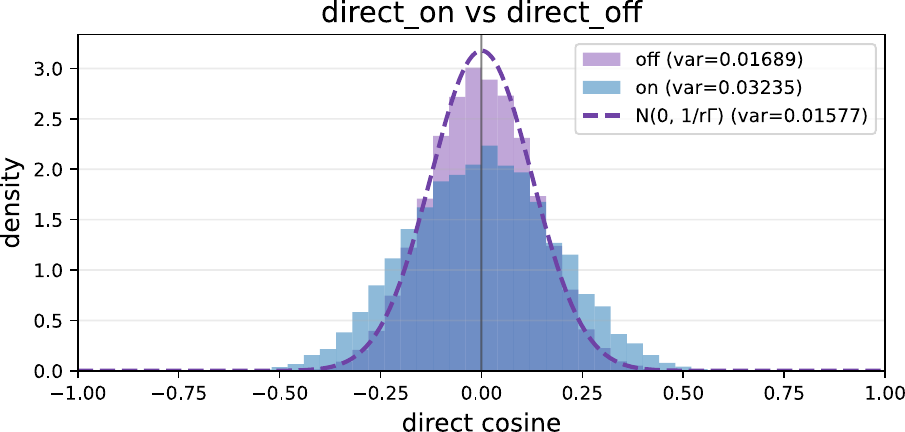}
    \caption{2nd layer of random model.}
    \label{sub:2nd-random}
  \end{subfigure}
  \hfill 
  \begin{subfigure}[b]{0.48\textwidth}
    \centering
    \includegraphics[width=\textwidth]{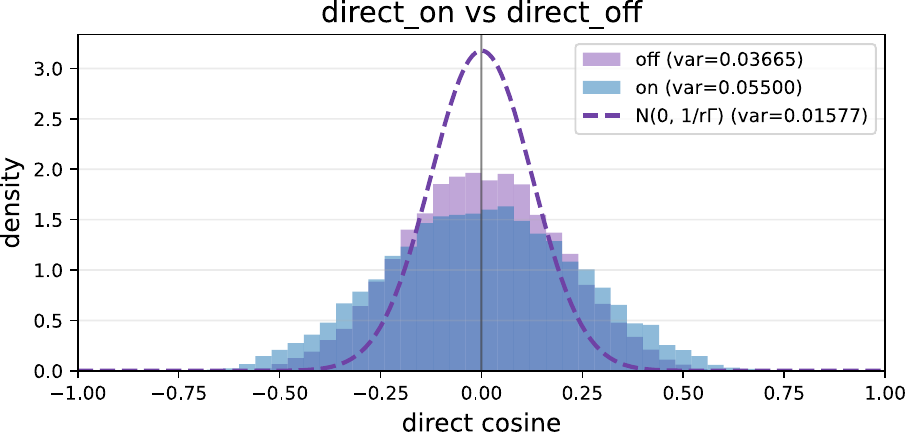}
    \caption{2nd layer of trained model.}
    \label{sub:2nd-trained}
  \end{subfigure}
  \vspace{2ex} 

  \begin{subfigure}[b]{0.48\textwidth}
    \centering
    \includegraphics[width=\textwidth]{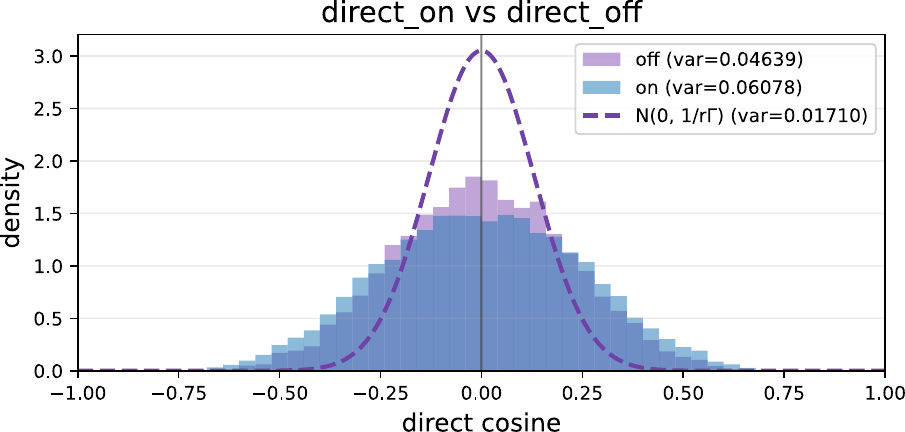}
    \caption{4th layer of random model.}
    \label{sub:4th-random}
  \end{subfigure}
  \hfill 
  \begin{subfigure}[b]{0.48\textwidth}
    \centering
    \includegraphics[width=\textwidth]{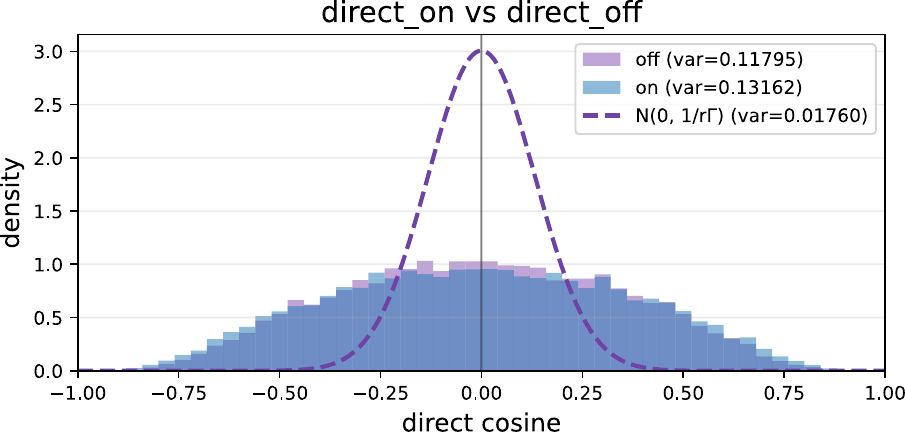}
    \caption{4th layer of trained model.}
    \label{sub:4th-trained}
  \end{subfigure}
  \caption{Direct cosine distributions for random and MNIST-trained models.}
  \label{fig:direct}
\end{figure}
We experimentally observed the direct cosine distributions for both initialized and MNIST-trained models with 784 inputs, 5 hidden layers of width 128, and 10 outputs. 
\Cref{fig:direct} summarizes the results, where $100$ \intersectionspaces are used for each neuron, giving 12,800 spaces per evaluated layer. 

The experiment for the 2nd layer of the initialized model shows a good fit to the random cosine assumption, as shown in \Cref{sub:2nd-random}. 
However, \Cref{sub:4th-random} shows that the fit rapidly degrades as the target layer becomes deeper. 
Moreover, it is not useful for the trained model (see \Cref{sub:2nd-trained,sub:4th-trained}). 



\subsection{\texorpdfstring{$\bfGamma$}{M}-Weighted Cosine Method}
Cosine similarity in the input space is strongly affected by the distortion of the local linear map $\bfGamma$, resulting in the direct cosine method not working in the deep layer. 
To mitigate this effect, we map the recovered decision-boundary normal $\bmv_\epsilon$ to the input space of the target layer. 

From the input space, the target layer can be reached only through the local linear map $\bfGamma$. For example, when $r_{\bfGamma}$ is small, the target-layer normal cannot be reconstructed exactly. Therefore, comparison must be performed in the image of $\bfGamma$, namely, in the space visible through the local linear map $\bfGamma$.
The projection matrix associated with $\bfGamma$ is
\[
P_{\bfGamma}:=\bfGamma(\bfGamma^\top\bfGamma)^+\bfGamma^\top
=\bfGamma\bfGamma^+=(\bfGamma^\top)^+\bfGamma^\top.
\]
Applying the pseudoinverse $(\bfGamma^\top)^+$ to the decision-boundary normal $\bmv_\epsilon$ gives
\[
\tilde \bmv_\epsilon:=(\bfGamma^\top)^+\bmv_\epsilon.
\]
This maps the input-space decision-boundary normal to the visible subspace of the target layer's input. Similarly, we project each signature $\alpha_j \bmw_j$ onto the same visible subspace:
\[
\alpha_j \tilde \bmw_j:=P_{\bfGamma} \alpha_j \bmw_j.
\]
The $\bfGamma$-weighted cosine method uses $z_{\epsilon}^{(\bfGamma)} = \cos(\alpha_t \tilde \bmw_t,\tilde \bmv_\epsilon)$, which removes the singular-value-dependent expansion and contraction induced by $\bfGamma^\top$ in the input space and measures cosine in the visible subspace of the target layer's input.

\begin{model}[$\bfGamma$-weighted cosine] \label{model:gamma}
  We assume that a $\bfGamma$-weighted cosine follows 
  \begin{align}
    z_{\mathrm{off}}^{(\bfGamma)} \overset{\mathrm{approx}}{\sim} \mathcal{N}\!\left(0,\frac{1}{r_{\bfGamma}}\right), && 
    z_{\mathrm{on}}^{(\bfGamma)} \overset{\mathrm{approx}}{\sim} \mathcal{N}\!\left(0,\frac{1}{r_{\bfGamma}} + \frac{1}{| \mathcal{J}_\mathrm{act} |} \right),
  \end{align} 
  where $\mathcal{J}_{\mathrm{act}}$ is the set of active neurons contributing to the off-side normal. 
\end{model}

The off-side model is simply derived from an assumption that $\alpha_t \tilde \bmw_t$ and $\tilde \bmv_{\mathrm{off}}$ are independent and isotropic in direction in the $r_{\bfGamma}$-dimensional visible subspace. 

In the on-side model, only the activation state of the target neuron changes. 
A $\tilde \bmw_t$ component is added to the off-side decision-boundary normal $\tilde \bmv_{\mathrm{off}}$. 
\[ 
\tilde \bmv_{\mathrm{on}}=\tilde \bmv_{\mathrm{off}}+\beta_t\tilde \bmw_t = \sum_{j \in \mathcal{J}_{\mathrm{act}}} \beta_j \tilde \bmw_j + \beta_t\tilde \bmw_t.
\]
Thus, the on-side cosine is
\[
z_{\mathrm{on}}^{(\bfGamma)} =\cos(\alpha_t \tilde \bmw_t,\tilde \bmv_{\mathrm{on}}) =\frac{\alpha_t \tilde \bmw_t^\top\tilde \bmv_{\mathrm{off}} + \alpha_t \tilde \bmw_t^\top\beta_t\tilde \bmw_t}{\|\tilde \bmw_t\|\,\|\tilde \bmv_{\mathrm{on}}\|}.
\]
If the target component is sufficiently small compared with the off-side background, we approximate $\|\tilde \bmv_{\mathrm{on}}\|\approx\|\tilde \bmv_{\mathrm{off}}\|$. Then,
\[
z_{\mathrm{on}}^{(\bfGamma)}
\approx z_{\mathrm{off}}^{(\bfGamma)}+
\frac{\alpha_t \beta_t\|\tilde \bmw_t\|}{\|\tilde \bmv_{\mathrm{off}}\|}
=z_{\mathrm{off}}^{(\bfGamma)}+
\frac{\alpha_t \beta_t\|\tilde \bmw_t\|}
{\left\|\sum_{j \in \mathcal{J}_{\mathrm{act}}}\beta_j \tilde \bmw_j\right\|}.
\]
In high dimensions with sufficiently many active components, the cross terms between distinct $\tilde \bmw_j$ tend to cancel on average. Therefore,
$\left\|\sum_{j \in \mathcal{J}_{\mathrm{act}}} \beta_j \tilde \bmw_j\right\|^2 \approx\sum_{j \in \mathcal{J}_{\mathrm{act}}} \beta_j^2\|\tilde \bmw_j \|^2$. 
We further assume that $\tilde \bmw_j$ comes from distributions of comparable scale independently of $j$. 
Under this assumption, the variance of $z_{\mathrm{on}}^{(\bfGamma)}$ typically becomes
\[
\operatorname{Var}\left(z_{\mathrm{on}}^{(\bfGamma)}\right)
\approx\frac{1}{r_{\bfGamma}}+
\frac{\beta_t^2}{\sum_{j \in \mathcal{J}_{\mathrm{act}}}\beta_j^2}
\approx\frac{1}{r_{\bfGamma}}+\frac{1}{|\mathcal{J}_{\mathrm{act}}|},
\]
with an assumption that $\beta_j$ is independently and identically distributed across \intersectionspaces and $j$ with a mean of zero. 
As a result, the on-side model is derived by assuming that $z_{\mathrm{on}}^{(\bfGamma)}$ follows a normal distribution. 

\subsubsection{Experiments.}

\begin{figure}[t]
  \centering
  \begin{subfigure}[b]{0.48\textwidth}
    \centering
    \includegraphics[width=\textwidth]{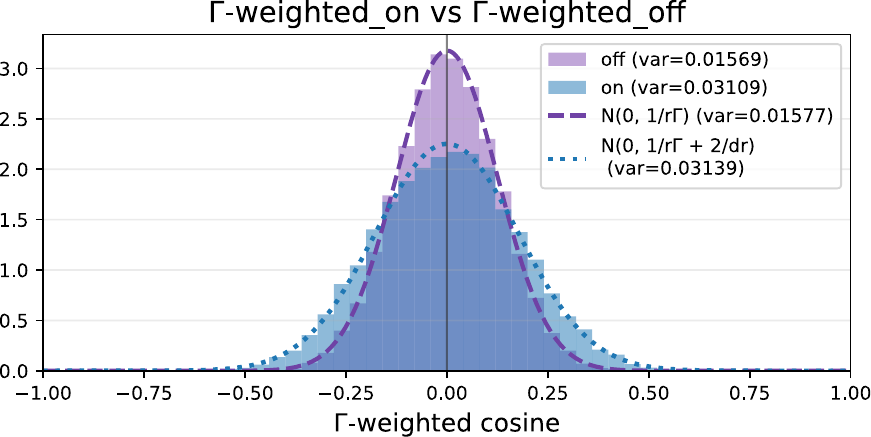}
    \caption{2nd layer of random model.}
    \label{sub:2nd-random-gamma}
  \end{subfigure}
  \hfill 
  \begin{subfigure}[b]{0.48\textwidth}
    \centering
    \includegraphics[width=\textwidth]{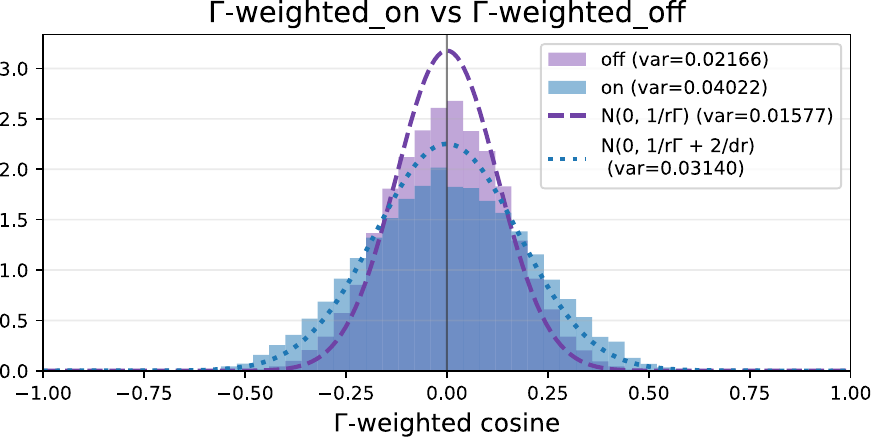}
    \caption{2nd layer of trained model.}
    \label{sub:2nd-trained-gamma}
  \end{subfigure}
  \vspace{2ex} 

  \begin{subfigure}[b]{0.48\textwidth}
    \centering
    \includegraphics[width=\textwidth]{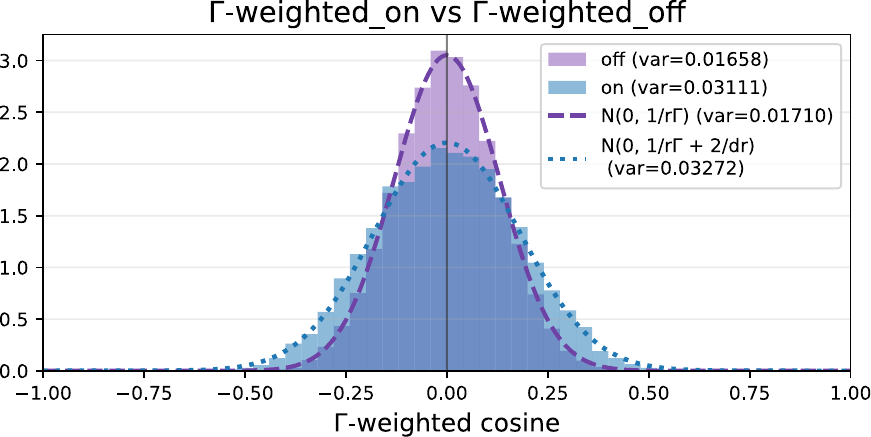}
    \caption{4th layer of random model.}
    \label{sub:4th-random-gamma}
  \end{subfigure}
  \hfill 
  \begin{subfigure}[b]{0.48\textwidth}
    \centering
    \includegraphics[width=\textwidth]{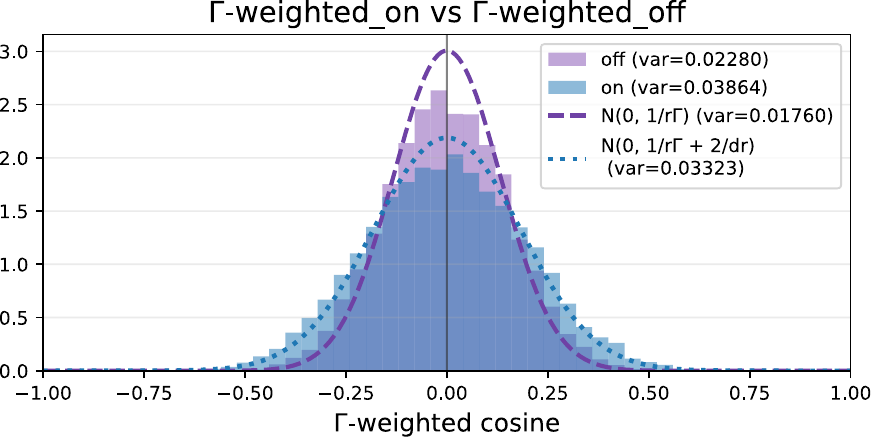}
    \caption{4th layer of trained model.}
    \label{sub:4th-trained-gamma}
  \end{subfigure}
  \caption{$\Gamma$-weighted cosine distributions for random and MNIST-trained models.}
  \label{fig:gamma}
\end{figure}

\Cref{model:gamma} is derived based on many assumptions. 
We therefore assess this model empirically, and \Cref{fig:gamma} summarizes the results.  
The target model and utilized \intersectionspaces are exactly the same as those in \Cref{fig:direct}. 
\Cref{model:gamma} fits the experimental results well in the initialized model. 
On the other hand, we still observe a gap in the trained model.

\subsection{Signature-Weighted Cosine Method}
\label{subsec:signature}

The weights of a trained model cannot be assumed to be random. Multiple neuron weights may be trained to concentrate in the same direction. If the projected target weight $\tilde \bmw_t$ is relatively aligned with the other $\tilde \bmw_j$, then, because $\tilde \bmv_{\mathrm{off}}=\sum_{j\in \mathcal{J}_{\mathrm{act}}}\beta_j\tilde \bmw_j$, the cosine similarity between $\tilde \bmv_{\mathrm{off}}$ and $\tilde \bmw_t$ has a large variance. As a result, distinguishing $\tilde \bmv_{\mathrm{on}}$ from $\tilde \bmv_{\mathrm{off}}$ becomes difficult.

To remove this anisotropy, we compare cosine similarities in a space whitened using the projected recovered signatures $\alpha_j \tilde \bmw_j$. 
This whitening suppresses contributions from dense directions and evaluates contributions from rare directions more strongly. 
Ideally, the competitor space should be formed by recovered signatures of active non-target neurons. 
However, the attacker cannot determine which neuron is active without knowing each sign. 
Therefore, the competitor space is formed by the recovered signatures of all non-target neurons instead. 
We assume that the distribution of directions of all non-target signatures approximately reflects the directional bias produced by real active competitors. 
We call this method the \emph{signature-weighted cosine method}. 



Again, we use $\alpha_j \tilde \bmw_j=P_{\bfGamma}\alpha_j \bmw_j$ and $\tilde \bmv_\epsilon=(\bfGamma^\top)^+\bmv_\epsilon$. 
Let $\mathcal{J}_{\mathrm{comp}}:=\{1,\ldots,d_\ell\}\setminus\{t\}$ be the index set of neurons in the same layer other than the target neuron. 
Let
\[
\tilde \bfC:=\sum_{j\in \mathcal{J}_{\mathrm{comp}}} (\alpha_j \tilde \bmw_j) (\alpha_j \tilde \bmw_j)^\top = 
\sum_{j\in \mathcal{J}_{\mathrm{comp}}} \tilde \bmw_j \tilde \bmw_j^\top.
\]
Note that the correction matrix $\tilde \bfC$ constructed from the recovered signatures is identical to that constructed from the true weights.
Then, the signature-weighted cosine is defined by
\[
z_\epsilon:=\cos_{\tilde \bfC}(\alpha_t \tilde \bmw_t,\tilde \bmv_\epsilon) :=\cos(\alpha_t \tilde \bfC^{-1/2} \tilde \bmw_t, \tilde \bfC^{-1/2} \tilde \bmv_\epsilon),
\]
where we consider the case $\operatorname{rank}(\tilde\bfC)=r_{\bfGamma}$ and interpret $\tilde\bfC^{-1/2}$ as the inverse square root restricted to $\operatorname{Im}(P_{\bfGamma})$.
Note that when it is rank-deficient, we can use another strategy, a \emph{projection-based method} in \Cref{subsec:correct}.

\begin{model}[Signature-weighted cosine] \label{model:signature}
  We assume that a signature-weighted cosine follows 
  \begin{align}
    z_{\mathrm{off}} \overset{\mathrm{approx}}{\sim} \mathcal{N}\!\left(0,\frac{1}{r_{\bfGamma}}\right), && 
    z_{\mathrm{on}}  \overset{\mathrm{approx}}{\sim} \mathcal{N}\!\left(0,\frac{1}{r_{\bfGamma}} + \frac{\kappa}{| \mathcal{J}_\mathrm{act} |} \right).
  \end{align} 
  where $\kappa = \frac{|\mathcal{J}_{\mathrm{comp}}|}{|\mathcal{J}_{\mathrm{comp}}|-r_{\bfGamma}-1}$ with $|\mathcal{J}_{\mathrm{comp}}| > r_\bfGamma + 1$. 
\end{model}

Applying $\tilde \bfC^{-1/2}$ is standard whitening that removes this anisotropy. 
It suppresses directions in which competitors are dense, while strengthening directions in which competitors are sparse.
As a result, $\tilde \bfC^{-1/2}\tilde \bmv_{\mathrm{off}}$ approaches an approximately isotropic vector in $r_{\bfGamma}$ dimensions. 


The purpose of the signature-weighted cosine method is to remove the direction-dependent bias of $\tilde \bmv_{\mathrm{off}}$. However, it also relatively emphasizes the target component added on the on side. This is because $\tilde \bfC$ is constructed only from competitors and does not include $\tilde \bmw_t$. 

In a local region where only the target ReLU switches,
\[
\tilde \bmv_{\mathrm{on}}=\tilde \bmv_{\mathrm{off}}+\beta_t\tilde \bmw_t,
\qquad
\tilde \bfC^{-1/2}\tilde \bmv_{\mathrm{on}}
=\tilde \bfC^{-1/2}\tilde \bmv_{\mathrm{off}}+\beta_t\tilde \bfC^{-1/2}\tilde \bmw_t.
\]
Thus,
\[
z_{\mathrm{on}}=
\frac{(\alpha_t \tilde \bfC^{-1/2}\tilde \bmw_t)^\top(\tilde \bfC^{-1/2}\tilde \bmv_{\mathrm{off}})+\alpha_t \beta_t\|\tilde \bfC^{-1/2}\tilde \bmw_t\|^2}
{\|\tilde \bfC^{-1/2}\tilde \bmw_t\|\,\|\tilde \bfC^{-1/2}\tilde \bmv_{\mathrm{on}}\|}.
\]
Assuming that the target component is sufficiently small relative to the off-side background so that $\|\tilde \bfC^{-1/2}\tilde \bmv_{\mathrm{on}}\|\approx\|\tilde \bfC^{-1/2}\tilde \bmv_{\mathrm{off}}\|$, we obtain
\[
z_{\mathrm{on}}
\approx z_{\mathrm{off}}+
\frac{\alpha_t \beta_t\|\tilde \bfC^{-1/2}\tilde \bmw_t\|}
{\|\tilde \bfC^{-1/2}\tilde \bmv_{\mathrm{off}}\|}.
\]
This additional term is amplified from 
$
\alpha_t \beta_t\|\tilde \bmw_t\|/\|\tilde \bmv_{\mathrm{off}}\|
$
in the $\bfGamma$-weighted cosine. 
As derived in \Cref{sec:on-side-amp}, the typical squared magnitude is estimated as
\[
\left( 
\frac{\alpha_t \beta_t\|\tilde \bfC^{-1/2}\tilde \bmw_t\|}
{\|\tilde \bfC^{-1/2}\tilde \bmv_{\mathrm{off}}\|}
\right)^2
\approx
\frac{|\mathcal{J}_{\mathrm{comp}}|}{|\mathcal{J}_{\mathrm{comp}}|-r_{\bfGamma}-1}
\frac{\beta_t^2}{\sum_{j\in \mathcal{J}_{\mathrm{act}}}\beta_j^2}
\approx
\frac{|\mathcal{J}_{\mathrm{comp}}|}{|\mathcal{J}_{\mathrm{comp}}|-r_{\bfGamma}-1}
\frac{1}{|\mathcal{J}_{\mathrm{act}}|}.
\]

\subsubsection{Experiments.}

\begin{figure}[t]
  \centering
  \begin{subfigure}[b]{0.48\textwidth}
    \centering
    \includegraphics[width=\textwidth]{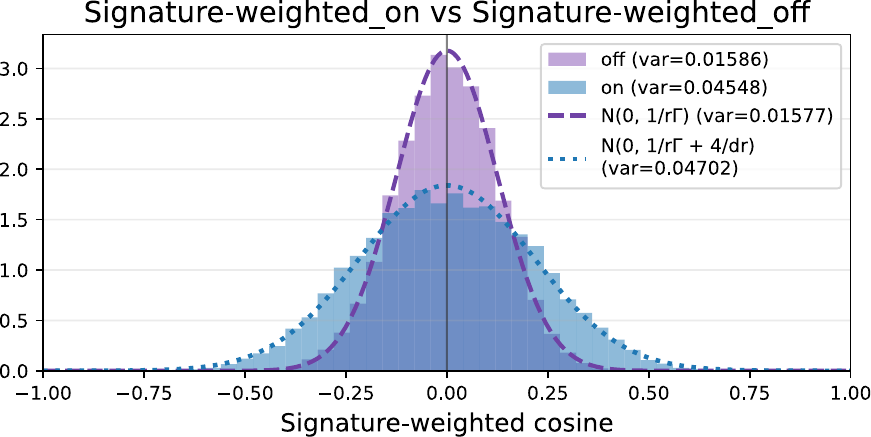}
    \caption{2nd layer of random model.}
    \label{sub:2nd-random-signature}
  \end{subfigure}
  \hfill 
  \begin{subfigure}[b]{0.48\textwidth}
    \centering
    \includegraphics[width=\textwidth]{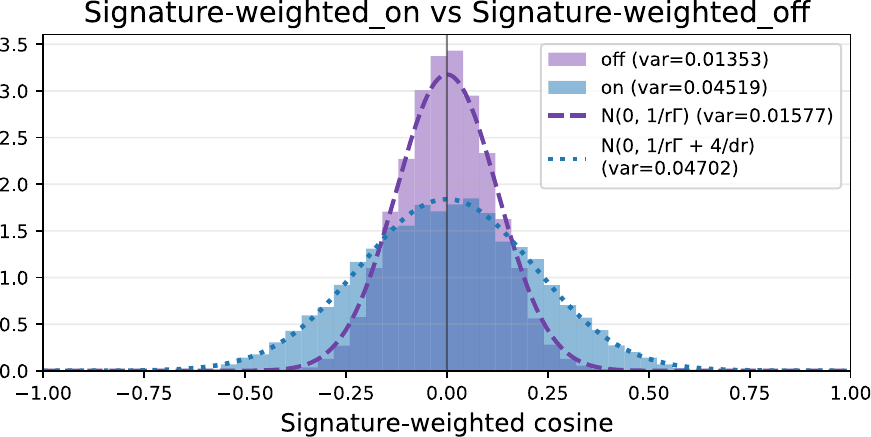}
    \caption{2nd layer of trained model.}
    \label{sub:2nd-trained-signature}
  \end{subfigure}
  \vspace{2ex} 

  \begin{subfigure}[b]{0.48\textwidth}
    \centering
    \includegraphics[width=\textwidth]{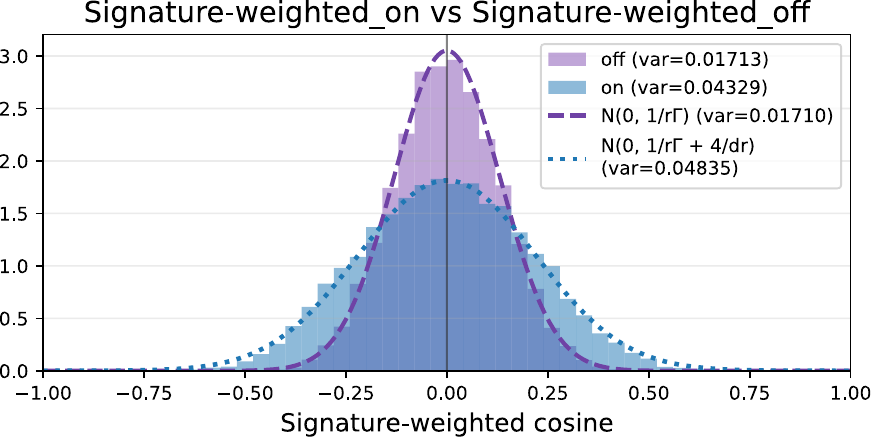}
    \caption{4th layer of random model.}
    \label{sub:4th-random-signature}
  \end{subfigure}
  \hfill 
  \begin{subfigure}[b]{0.48\textwidth}
    \centering
    \includegraphics[width=\textwidth]{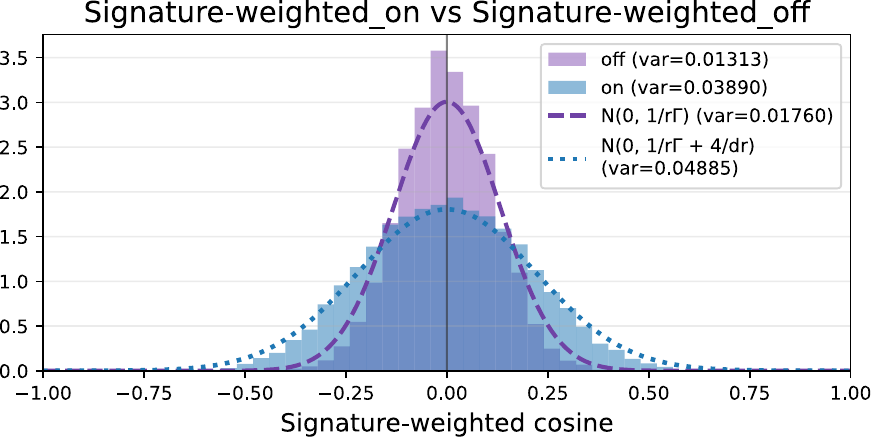}
    \caption{4th layer of trained model.}
    \label{sub:4th-trained-signature}
  \end{subfigure}
  \caption{Signature-weighted cosine distributions for random and MNIST-trained models.}
  \label{fig:signature}
\end{figure}

\Cref{fig:signature} summarizes the results of the signature-weighted cosine method. 
Again, the target model and utilized \intersectionspaces are exactly the same as those in \Cref{fig:direct,fig:gamma}. 
Using the signature-weighted cosine method produces results closer to the expected model than the $\bfGamma$-weighted cosine method.
Besides, we can observe that the cosine distributions for the on side yield larger variance.

\subsection{Algorithm and Estimating the Number of Intersection Spaces}

\begin{algorithm}[t]
\caption{\textsc{SignatureWeightedSignRecovery}}
\label{alg:signature-weighted-sign-recovery}
\begin{algorithmic}[1]
\Require $N$ \intersectionspaces; target $\alpha_t \bmw_t$; competitors $\{\alpha_j \bmw_j\}_{j\ne t}$; recovered preceding-layer weights and biases
\Ensure the sign information for the target neuron
\State $\mathcal{D}\gets\emptyset$
\ForAll{given \intersectionspace $(\bms, \bmv, \bmv')$}
    \State Obtain adjacent boundary points $\bmx$ and $\bmx'$.
    \State $\bfGamma\gets{\bf\Gamma}^{(\ell-1)}_{\bms}$.
    \If{$\alpha_t \bmw_t^\top\bfGamma(\bmx-\bms)<0$}
        \State $(\bmv,\bmv') \gets (\bmv',\bmv)$.
    \EndIf
    \State $(\tilde{\bmv}, \tilde{\bmv}')  \gets \left( (\bfGamma^\top)^+ \bmv, (\bfGamma^\top)^+ \bmv' \right)$.
    \State $\alpha_j \tilde{\bmw}_j\gets P_{\bfGamma} (\alpha_j \bmw_j)$ for $j=1,\ldots,d_\ell$.
    \State $\tilde{\bfC}\gets\sum_{j\in\mathcal{J}_{\mathrm{comp}}} (\alpha_j \tilde{\bmw}_j)(\alpha_j \tilde{\bmw}_j)^\top$.
    \State $(z, z') \gets \left( \cos_{\tilde{\bfC}}(\alpha_t \tilde{\bmw}_t,\tilde{\bmv}), \cos_{\tilde{\bfC}}(\alpha_t \tilde{\bmw}_t,\tilde{\bmv'}) \right) $.
    \State Append $z^2-z'^2$ to $\mathcal{D}$.
\EndFor
\State $T\gets \frac{\sqrt{N}\,\overline{\mathcal{D}}}{s_{\mathcal{D}}}$, where $\overline{\mathcal{D}}$ and $s_{\mathcal{D}}$ are the sample mean and standard deviation of $\mathcal{D}$.
\If{$T<0$}
    \Return $\alpha_t = -1$.
\EndIf
\State \Return $\alpha_t = 1$.
\end{algorithmic}
\end{algorithm}
\Cref{alg:signature-weighted-sign-recovery} shows the pseudocode for sign recovery. 
Each \intersectionspace yields two values, $z_i$ and $z_i'$. The value aggregated for sign classification is $D_i:=z_i^2 - {z'}_i^2$.
This score directly measures the difference in their variances when their means can be approximated as zero.
To estimate the required number of \intersectionspaces, we regard both sides as zero-mean normal distributions. 
Under an approximation that ignores covariance,
\[
\E[D_i]=\operatorname{Var}(z)-\operatorname{Var}(z'),
\qquad
\operatorname{Var}(D_i)\simeq2\operatorname{Var}(z)^2+2\operatorname{Var}(z')^2.
\]
Therefore, the expected $T$ value for $N$ samples is
\[
T \simeq \sqrt{N}\,
\frac{\operatorname{Var}(z)-\operatorname{Var}(z')}{\sqrt{2\operatorname{Var}(z)^2+2\operatorname{Var}(z')^2}}.
\]
Thus, when $T>0$, $\bmv$ is the on-side. When $T < 0$, $\bmv'$ is the on side.

For the signature-weighted cosine method, the additional on-side term is about 
$\frac{|\mathcal{J}_{\mathrm{comp}}|}{|\mathcal{J}_{\mathrm{comp}}|-r_{\bfGamma}-1}\frac{1}{|\mathcal{J}_{\mathrm{act}}|} \approx 4/d_\ell$ when $|\mathcal{J}_{\mathrm{comp}}| = d_\ell - 1$ and $|\mathcal{J}_{\mathrm{act}}| \approx r_\bfGamma \approx d_\ell/2$. 
\[
\operatorname{Var}(z_{\mathrm{off}})\simeq\frac{1}{r_{\bfGamma}}\simeq\frac{2}{d_\ell},
\qquad
\operatorname{Var}(z_{\mathrm{on}})\simeq\frac{2}{d_\ell}+\frac{4}{d_\ell}=\frac{6}{d_\ell}.
\]
Then, $\E[D_i] \simeq 4/d_\ell$ and $\operatorname{Var}(D_i)\simeq 80/d_\ell^2$, and $N\gtrsim45$ yields $|T| \ge 3$ under the independent normal approximation assumption.

\subsection{Correcting Unreliable Signs} 
\label{subsec:correct}

The signature-weighted cosine method is based on empirical statistical bias. 
Since the underlying assumption does not always hold in trained models, we need to verify the recovered signs and correct the overall result if necessary. 
For this purpose, we introduce two methods.

\subsubsection{Active-Weighted Cosine Method.}
As mentioned in \Cref{subsec:signature}, ideally, the competitor space should be formed by active weights $\mathcal{J}_{\mathrm{act}}$. 
Actually, the signature-weighted cosine method also whitens the weights of neurons that are actually inactive and do not contribute to $\tilde \bmv_{\mathrm{off}}$. 
Therefore, it is reasonable to fix highly reliable neuron signs, e.g., $|t| \ge 3$ in the signature-weighted cosine method, and remove these weights from the competitor space if they are inactive in each \intersectionspace. 
Let $\mathcal{J}_\mathrm{cand} := \mathcal{J}_{\mathrm{comp}} \setminus \mathcal{J}_{\mathrm{inactive}}$, where $\mathcal{J}_{\mathrm{inactive}}$ contains indices whose neurons are identified as inactive according to highly reliable signs. Then, 
\[
\tilde \bfC_{\mathrm{cand}}:=\sum_{j\in \mathcal{J}_{\mathrm{cand}}} (\alpha_j \tilde \bmw_j)(\alpha_j \tilde \bmw_j)^\top
=\sum_{j\in \mathcal{J}_{\mathrm{cand}}} \tilde \bmw_j \tilde \bmw_j^\top.
\]
When $\tilde\bfC_{\mathrm{cand}}$ is invertible on $\operatorname{Im}(P_{\bfGamma})$, we use $\cos_{\tilde\bfC_{\mathrm{cand}}}(\alpha_t \tilde \bmw_t, \tilde \bmv_\epsilon)$.
To distinguish between the terms, we call this method the \emph{active-weighted cosine method}. 
The rank-deficient case is handled by the projection-based method described below.


\subsubsection{Projection-Based Method.}
After recovering many signs, we may obtain an \intersectionspace satisfying $\operatorname{rank}(\tilde \bfC_{\mathrm{cand}})<r_{\bfGamma}$. 
Then, we can use a powerful analytic method. 

\begin{proposition} \label{prop:low-rank}
  Suppose that $\mathcal J_{\mathrm{act}} \subseteq \mathcal J_{\mathrm{cand}}$ and $\operatorname{rank}(\tilde \bfC_{\mathrm{cand}})<r_{\bfGamma}$. 
  Then, the off-side normal $\tilde \bmv_{\mathrm{off}}$ lies in the competitor span $\mathrm{Im}(\tilde \bfC_{\mathrm{cand}})$. Unless $\tilde \bmw_{t}$ happens to lie entirely in the competitor span, projection onto the competitor span leaves the off-side normal unchanged and changes the on-side normal. Thus, the sign can be determined directly from a single \intersectionspace usually.
\end{proposition}
\Cref{prop:low-rank} implies that the sign is easily recovered once $\operatorname{rank}(\tilde \bfC_{\mathrm{cand}})<r_{\bfGamma}$ holds. 
Let's consider the model using $784\--128\--\cdots$ and try to recover the signs in the 1st layer. 
Then, $\operatorname{rank}(\tilde \bfC_{\mathrm{cand}}) \le 127$ and $r_{\bfGamma} = 784$, and this condition always holds. 
Thus, we can recover all the signs in the first layer. 

For subsequent layers, this condition does not hold initially because we cannot identify which neurons are active. 
However, after recovering and fixing the signs of almost all neurons, some \intersectionspaces could satisfy this condition. 
Once we find it, we can determine all signs. 
Specifically, all signs identified as inactive neurons are correct because $\tilde \bmv_\mathrm{off}$ lies in the competitor span. 
For neurons identified as active, we remove such a neuron from the competitor span one by one and verify whether $\tilde \bmv_\mathrm{off}$ still lies in the competitor span. 
In this way, we can determine the signs of all neurons. 


In a noise-free environment, this projection-based method can reliably recover each sign, except in negligible cases such as when $\beta_j$ is exactly 0 or when a weight can be fully described by other weights.
However, in practice, end-to-end extraction always introduces non-negligible computational noise, which means this method is not fully deterministic. 
This is one of the reasons that the actual end-to-end extraction is significantly harder than sign recovery alone. 
For details, please refer to \Cref{sec:e2e}.

\subsection{Comparison with the Existing Method} \label{subsec:comparison}


\begin{table}[t]
\centering
\caption{Sign-recovery results for an MNIST-trained $784$-$128^{(5)}$-$10$ model from 50 \intersectionspaces per evaluated neuron. Numbers labeled by HC denote high-confidence results, where $|t| \ge 3$ for the signature-weighted cosine method and $CL \ge 99.7\%$ for the boundary-walking method. Here, HC under Wrong denotes a high-confidence error, namely, a case in which the incorrect sign direction is selected with high confidence.}
\label{tab:sign-comparison}
\begin{tabular}{c|c|cc|cc}
\toprule
& & \multicolumn{2}{c|}{Signature-weighted cosine}
& \multicolumn{2}{c}{Boundary-walking} \\
Layer
& Evaluated
& Correct (HC)
& Wrong (HC)
& Correct (HC)
& Wrong (HC) \\
\midrule
2 & 128 & 127 (119) & 1 (0) & 105 (0) & 23 (0) \\
3 & 128 & 128 (110) & 0 (0) & 85 (0)  & 43 (0) \\
4 & 128 & 128 (120) & 0 (0) & 80 (0)  & 48 (0) \\
5 & 126 & 123 (107) & 3 (0) & 67 (0)  & 59 (0) \\
\bottomrule
\end{tabular}
\end{table}

We compare the accuracy of sign recovery by the signature-weighted cosine method and the existing boundary-walking method using the same \intersectionspaces. 
The experiment is conducted in a setting where \intersectionspaces, target-layer signatures, past-layer weights, and past-layer biases are given without noise. 
Under such a setting, our method does not require additional dedicated queries. 
In contrast, the boundary-walk method requires dedicated queries, and we evaluated it using a white-box implementation, as in prior work. 
The target model is an MNIST-trained $784$-$128^{(5)}$-$10$ model. 
We used 50 \intersectionspaces for each evaluated neuron, excluding persistent and dead neurons. 

\Cref{tab:sign-comparison} summarizes the sign-recovery result for each layer of this target model. 
The boundary-walking method did not yield any high-confidence results, indicating that 50 \intersectionspaces are insufficient. 
This is not surprising, as prior work has already reported that about $10^2$-$10^3$ \intersectionspaces are required~\cite{DBLP:conf/eurocrypt/CarliniCHRS25}. 
In contrast, the signature-weighted cosine method recovers the signs of almost all neurons with a high confidence level. 

Although the signature-weighted cosine method performs much better than the existing method, opposite signs are wrongly recovered in a few neurons, e.g., one and three neurons in the 2nd and 5th layers. 
As discussed in \Cref{subsec:correct}, the projection-based method can identify or correct additional signs when the required span conditions hold. 
In fact, we could verify and correct all signs of this model with the projection-based method. 
If the projection-based method fails to verify, we can proceed to the active-weighted cosine method, in which inactive neurons whose signs are recovered with high reliability are excluded from the competitor. 

%% file: 04_e2e.tex
\section{End-to-End Extraction} \label{sec:e2e}

This section addresses the challenges of end-to-end hard-label model extraction and presents methods for overcoming them.
Recovering parameters in deeper layers introduces difficulties that do not arise in shallow layers: numerical errors in previously recovered weights accumulate, and activation patterns in subsequent layers exhibit limited variation.
These issues also affect the sign-recovery algorithms presented in \cref{sec:sign}, which must tolerate such errors to support end-to-end extraction.
We first describe the challenges in signature recovery and sign recovery and present our solutions.
We then evaluate the resulting end-to-end attack on models with four or six hidden layers of width 16.

We do not evaluate recovery of persistent neurons, which remain active for every input, because none were identified in our experiments.
Existing cross-layer extraction techniques~\cite{DBLP:conf/crypto/ItoMT26} could accommodate such neurons, but we do not evaluate this extension.

\subsection{Improving Signature Recovery}

We address three challenges in signature recovery for end-to-end extraction.
(i) Some weights are difficult to recover because the corresponding neurons in the preceding layer are rarely or never active on the target neuron's activation boundary; these are called \emph{query-intensive weights} and \emph{unreachable weights}.
(ii) \intersectionspaces associated with neurons in subsequent layers can pass the consistency check for the target layer, producing \emph{noise components}.
(iii) \intersectionspaces with shared activation patterns in subsequent layers can cause a decision-boundary normal to be recovered as a signature.
Issues (i) and (ii) also arise in soft-label extraction~\cite{DBLP:conf/eurocrypt/LiuSELBP26}, whereas (iii) is specific to hard labels.

\begin{figure}[t]
    \centering
    \begin{subfigure}[t]{0.49\linewidth}
        \centering
        \includegraphics[width=\linewidth]{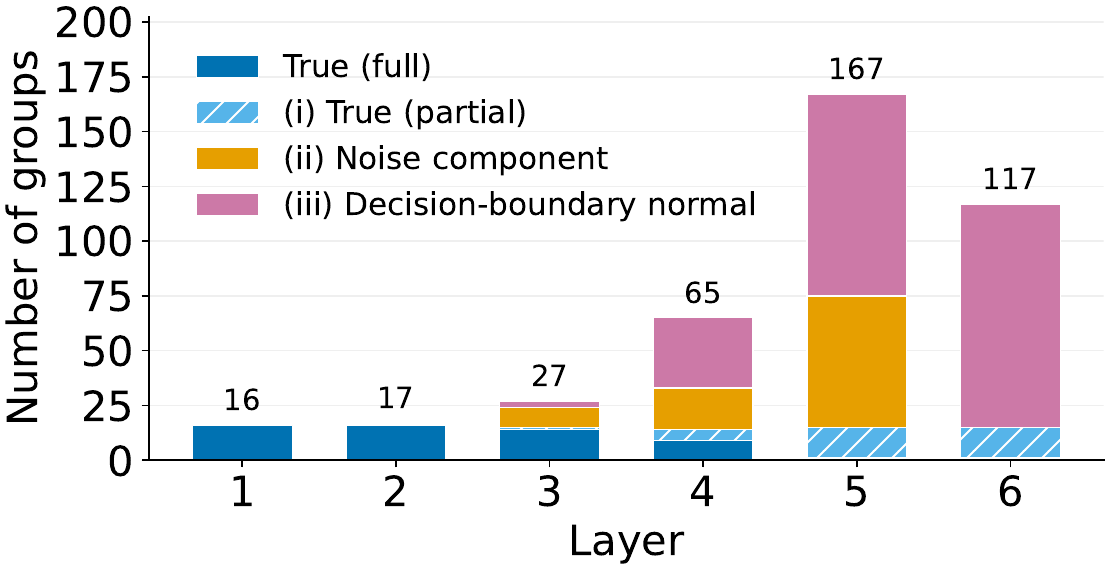}
        \caption{MNIST}
    \end{subfigure}\hfill
    \begin{subfigure}[t]{0.49\linewidth}
        \centering
        \includegraphics[width=\linewidth]{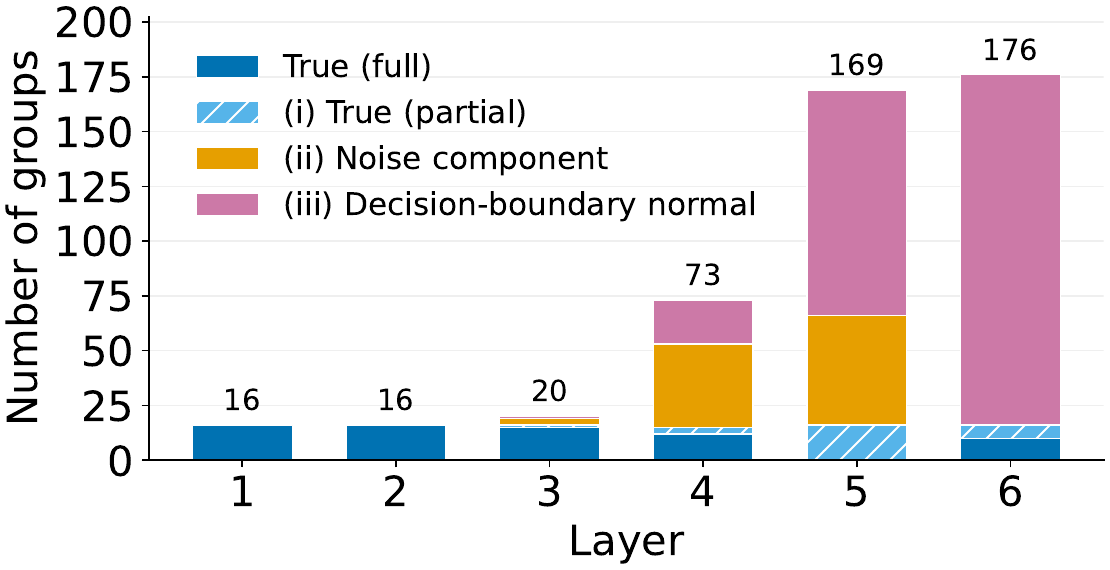}
        \caption{FMNIST}
    \end{subfigure}
    \caption{Groups passing the consistency check by layer, using 20,000 \intersectionspaces per trained model with six hidden layers of width 16.}
    \label{fig:e2e-consistent-groups}
\end{figure}

\Cref{fig:e2e-consistent-groups} shows the breakdown of groups passing the consistency check among 20,000 \intersectionspaces collected from each of two models with six hidden layers of width 16, trained on MNIST and FMNIST, respectively.
True (full) and True (partial) denote groups corresponding to genuine signatures for which all coordinates or only a subset of coordinates can be recovered, respectively.
True (partial) is associated with issue (i), while Noise component and Decision-boundary normal correspond to issues (ii) and (iii), respectively.
We identified True (full) and True (partial) groups through white-box analysis using the ground-truth weights of each layer.
Their combined count can fall below the layer width of 16 because 20,000 \intersectionspaces may be insufficient or because some neurons are dead.
The end-to-end experiments in \cref{subsec:e2e-experiments} use 400,000 \intersectionspaces for each model with six hidden layers.
As the figure illustrates, filtering out groups arising from issues (ii) and (iii) should essentially leave only those corresponding to genuine signatures of the target layer.
The missing coordinates of True (partial) signatures must then be recovered.
We describe each issue and our solution under the corresponding heading (i)--(iii) below.
When discussing signature recovery, we use $\bmw$ for a recovered signature candidate whose sign has yet to be determined, omitting the explicit factor $\alpha$.

\subsubsection{(i) Query-Intensive and Unreachable Weights.}

As observed in~\cite{DBLP:conf/eurocrypt/LiuSELBP26}, dependencies between the activation patterns of neurons in adjacent layers can make certain weights difficult to recover.
The same issue arises in the hard-label setting.
Consider recovering the signature of the $k$th neuron in layer $\ell$, whose weight vector is $\bmw_k^{(\ell)}$.
If the $j$th neuron in layer $\ell-1$ is usually or always inactive on the target neuron's activation boundary, the $j$th input coordinate to layer $\ell$ is correspondingly usually or always zero at the target neuron's \intersectionpoints.
Points at which this coordinate is zero do not provide information about the missing signature coordinate $w_{k,j}^{(\ell)}$.

In~\cite{DBLP:conf/eurocrypt/LiuSELBP26}, a weight is called \emph{query-intensive} if the preceding neuron is active with very low probability on the target neuron's activation boundary, requiring a large number of queries for recovery.
An effective weight is called \emph{unreachable} if the preceding neuron is never active on the target activation boundary, yet the two neurons can be active together away from that boundary. A weight is non-effective if the two neurons cannot be active together.
Effective query-intensive and unreachable weights cannot simply be ignored: the preceding neuron need not be dead and may be active when the target neuron is also active.
Indeed, the experiments in~\cite{DBLP:conf/eurocrypt/LiuSELBP26} show that unreachable weights reduce layer coverage, particularly in deeper layers.

We use two approaches to address this issue.
The first adaptively queries the model to search for \intersectionpoints that reveal the missing coordinate.
It applies when such points exist but are difficult to find, as in the case of query-intensive weights.
The second uses cross-layer extraction~\cite{DBLP:conf/crypto/ItoMT26} to recover the missing coordinate from activation boundaries in the next layer.
This enables recovery even without a suitable \intersectionpoint for the target neuron.

\paragraph{Searching for informative intersection points.}
Starting from an existing \intersectionpoint of the target neuron, we move along the intersection of its activation boundary and the decision boundary toward a region where the preceding neuron associated with the missing coordinate becomes active.
We then perturb the input into that region and search nearby for a new \intersectionpoint that reveals the missing coordinate.
The detailed search procedure, including direction updates and restarts, is given in \cref{app:informative-intersection-search}.

\paragraph{Cross-layer extraction.}
The first approach cannot recover unreachable weights, so we instead use cross-layer extraction~\cite{DBLP:conf/crypto/ItoMT26}.
This technique was originally proposed to recover the weights of \emph{persistent neurons}, which remain active for every input.
For such neurons, ReLU acts as the identity, allowing their weights to be combined with those of the next layer.
Their combined contributions can then be estimated using activation boundaries in that layer.

We apply this principle to unreachable weights; see \cref{app:cross-layer-unreachable}.

\subsubsection{(ii) Spurious Recovery of Subsequent-Layer Signatures.}

In soft-label extraction, incorrectly grouping critical points associated with subsequent-layer neurons produces \emph{noise components}~\cite{DBLP:conf/eurocrypt/LiuSELBP26}.
The same issue arises in consistency checks on \intersectionspaces in the hard-label setting.

Suppose that the attacker is performing consistency checks for layer $\ell$.
Let $\bms_1$ and $\bms_2$ be two \intersectionpoints associated with the $k$th neuron in layer $\ell+1$, whose weight vector and bias are $\bmw_k^{(\ell+1)}$ and $b_k^{(\ell+1)}$.
Suppose that layer $\ell$ has the same activation pattern at both points, represented by the diagonal matrix $\bfD^{(\ell)}$.
Let $\bmh_i^{(\ell-1)}=\bmh^{(\ell-1)}(\bms_i)$ be the input to layer $\ell$ at $\bms_i$, for $i\in\{1,2\}$.
Since both points lie on the same next-layer activation boundary,
\begin{align}
    \begin{cases}
        \bmw_k^{(\ell+1)\top}\bfD^{(\ell)}\bfW^{(\ell)}\bmh_1^{(\ell-1)}
        +\bmw_k^{(\ell+1)\top}\bfD^{(\ell)}\bmb^{(\ell)}+b_k^{(\ell+1)}=0,\\
        \bmw_k^{(\ell+1)\top}\bfD^{(\ell)}\bfW^{(\ell)}\bmh_2^{(\ell-1)}
        +\bmw_k^{(\ell+1)\top}\bfD^{(\ell)}\bmb^{(\ell)}+b_k^{(\ell+1)}=0.
    \end{cases}
\end{align}
Define $\tilde{\bmw}_k^{(\ell+1)} =\bfW^{(\ell)\top}\bfD^{(\ell)\top}\bmw_k^{(\ell+1)}, \tilde{b}_k^{(\ell+1)} =\bmw_k^{(\ell+1)\top}\bfD^{(\ell)}\bmb^{(\ell)}+b_k^{(\ell+1)}$.
Then both points satisfy $\tilde{\bmw}_k^{(\ell+1)\top}\bmh_i^{(\ell-1)} +\tilde{b}_k^{(\ell+1)}=0,~ i\in\{1,2\}$. This constraint holds not only at the representative points but throughout each corresponding \intersectionspace after applying the local affine map to the input space of layer $\ell$.
When $\tilde{\bmw}_k^{(\ell+1)}$ is nonzero, the mapped spaces lie in a common hyperplane and can pass the consistency check for layer $\ell$.

The same issue can arise even when the $j$th neuron in layer $\ell$ changes its activation state across the collected \intersectionpoints.
Suppose that each point corresponds to the $k$th neuron in layer $\ell+1$, while the state of the $j$th neuron in layer $\ell$ is fixed in a neighborhood of each point.
The class pair defining the decision boundary and the activation patterns of the other neurons from layer $\ell$ onward, excluding the boundary neuron $k$ in layer $\ell+1$, are shared, whereas activation patterns in preceding layers may vary freely.
The corresponding \intersectionspaces can then pass the consistency check even when the $j$th neuron in layer $\ell$ has different activation states.
Its output appears linearly in both the next-layer neuron's activation-boundary equation and the decision-boundary equation, so its contribution can be eliminated by taking a linear combination of the two equations.
Here, the \intersectionspaces for both activation states, mapped to the input space of layer $\ell$, lie in a common hyperplane.
\Cref{app:noise-component-derivation} derives this shared constraint, including the decision-boundary coefficients obtained from the network formed by subsequent layers.
We also call this a \emph{noise component}, since it results from accepting subsequent-layer \intersectionspaces as those of a target-layer neuron.

\paragraph{Solution.}
To filter out noise components in both cases, we check whether each recovered candidate is an actual target-layer activation boundary.
In the soft-label setting, noise components can be rejected by checking whether the activation boundary bends~\cite{DBLP:conf/eurocrypt/LiuSELBP26}.
This test cannot be applied directly in the hard-label setting, where only class labels are observable.
Instead, we search in other ReLU cells for intersections between the candidate activation boundary and the decision boundary, and check whether the latter bends at these intersections.
A noise component may define a hyperplane containing the mapped \intersectionspaces without corresponding to an actual activation boundary of the target layer.
Such a candidate need not induce a bend in the decision boundary at a new intersection, whereas a genuine activation boundary is expected to do so.
We use this distinction to validate candidate signatures empirically.

We start from collected \intersectionpoints that were not used to recover the candidate and move along an adjacent decision-boundary toward the candidate activation boundary.
We prioritize starting points with shorter predicted travel distances to reduce the likelihood of crossing other ReLU cells.
\Cref{app:candidate-intersection-search} gives the detailed procedure and the displacement calculation.

\subsubsection{(iii) Spurious Recovery of Decision-Boundary Normals.}

Limited variation in activation patterns in deeper layers underlies problems such as the noise components reported in~\cite{DBLP:conf/eurocrypt/LiuSELBP26}.
Even random inputs may produce only a limited set of activation patterns in subsequent layers.
Viewed from the input space of an intermediate layer, fewer ReLU cells are encountered, and activation boundaries subdivide the decision boundary less frequently.
As a result, multiple \intersectionspaces, mapped to that input space, are more likely to lie in a shared decision-boundary hyperplane.
They can therefore pass the consistency check, yielding the hyperplane normal and offset as a spurious signature and bias.

Let $\mathcal{S}_1$ and $\mathcal{S}_2$ be two \intersectionspaces whose images under the respective local affine maps to the input space of layer $\ell$ lie in a shared decision-boundary hyperplane.
Let $\bms_1\in\mathcal{S}_1$ and $\bms_2\in\mathcal{S}_2$ be their representative \intersectionpoints.
Let $\bmv^{(\ell)}$ and $b^{(\ell)}$ be the normal and offset of this hyperplane, respectively.
Then $\bmv^{(\ell)\top}\bmh^{(\ell-1)}(\bms_i)+b^{(\ell)}=0, ~ i\in\{1,2\}$.
This shared hyperplane can cause $\bmv^{(\ell)}$ to be incorrectly recovered as a signature.

These points may nevertheless appear adjacent to different hyperplanes in the model's input space.
Let $\bfGamma_1^{(\ell-1)}$ and $\bfGamma_2^{(\ell-1)}$ be the forward local linear matrices through layer $\ell-1$ at $\bms_1$ and $\bms_2$, respectively.
The corresponding input-space normals are $\bmv_1^{(1)}=\bfGamma_1^{(\ell-1)\top}\bmv^{(\ell)}, \bmv_2^{(1)}=\bfGamma_2^{(\ell-1)\top}\bmv^{(\ell)}$.
Activation patterns in preceding layers typically differ between the two points, yielding different local linear maps and potentially different input-space normals.
Thus, the shared decision boundary may not be apparent in the model's input space.

To detect such spurious signatures, we pull the recovered signature back to the input space at the representative point of each \intersectionspace in a consistent group and compare it with the two adjacent decision-boundary normals.
Let $\bms_1,\ldots,\bms_m$ be the intersection points of the spaces in the group, let $(\bmv_i,\bmv_i')$ be the pair of normals associated with $\bms_i$, and let $\bmw$ be the recovered signature.
Let $\bfGamma_i$ be the forward local linear matrix to the target layer's input at $\bms_i$.
If $\bmw$ is actually a shared decision-boundary normal in that space, its pullback should be parallel to one of the observed normals at every point: $\bmv_i\parallel\bfGamma_i^\top\bmw ~\text{or}~ \bmv_i'\parallel\bfGamma_i^\top\bmw,~ i\in[m]$.

For numerical robustness, we measure the fraction of aligned points in each group.
For each point, we compute the maximum absolute cosine similarity $c_i=\max\bigl( |\cos(\bfGamma_i^\top\bmw,\bmv_i)|, |\cos(\bfGamma_i^\top\bmw,\bmv_i')| \bigr)$.
We reject a group as a suspected decision-boundary component if $\frac{\#\{i\in[m]\mid c_i<\rho\}}{m}<\tau$. We use $\rho=0.97$ and $\tau=0.01$ throughout, as these gave the best experimental results.

\subsection{Stabilizing Sign Recovery}

In \cref{sec:sign}, we demonstrated that most signs can be recovered from a small number of \intersectionspaces when accurate signatures and preceding-layer parameters are available.
In end-to-end extraction, numerical errors and incomplete recovery make reliable sign recovery more challenging.
We therefore use a two-stage procedure that empirically improves robustness: first, we apply the methods of \cref{sec:sign} to assign signs in order of confidence, incorporating each assignment into the remaining estimates; then, we check the assignments against the collected decision-boundary normals and correct them when necessary.

\paragraph{Step 1: Provisional Sign Assignment.}
For each neuron with an unassigned sign, we apply a projection- or cosine-based method from \cref{sec:sign} to its \intersectionspaces.
Following \cref{sec:sign}, let $\mathcal{J}_{\mathrm{comp}}$ be the index set of recovered non-target neurons and $\mathcal{J}_{\mathrm{inactive}}$ the subset classified as inactive at the current point under the assigned signs.
We form $\mathcal{J}_{\mathrm{cand}}=\mathcal{J}_{\mathrm{comp}}\setminus\mathcal{J}_{\mathrm{inactive}}$ and compare the dimension of its signature span, pulled back to the input space, with the rank of the local linear map.
If the former is smaller than the latter and exactly one normal lies in the competitor span, the projection-based method estimates that normal as the off-side normal.
This can hold without known signs, often in the first layer and sometimes in the second.

When the projection-based method cannot determine any additional signs, we turn to cosine-based estimation.
If no signs are known, $\mathcal{J}_{\mathrm{inactive}}=\varnothing$, so we use the signature-weighted cosine method with $\mathcal{J}_{\mathrm{cand}}=\mathcal{J}_{\mathrm{comp}}$.
Once some signs are known, we use the active-weighted cosine method with the updated $\mathcal{J}_{\mathrm{cand}}$.
We evaluate signs using the statistic $T$ and the sign decision rule in \cref{alg:signature-weighted-sign-recovery}.
We process the \intersectionspaces for each neuron sequentially and fix a neuron's sign as soon as it satisfies $|T|\geq3$.
Each time a single sign is fixed, we update $\mathcal{J}_{\mathrm{inactive}}$ and $\mathcal{J}_{\mathrm{cand}}$ at each point and immediately return to the projection-based method for the remaining neurons.
When that method can no longer determine additional signs, we resume cosine-based estimation with the updated competitors.
If no remaining neuron reaches the threshold using the available \intersectionspaces, we provisionally assign the remaining signs according to their respective $T$ statistics.
Thus, we prioritize reliable assignments, then check the complete assignment in Step 2.

\paragraph{Step 2: Sign Correction Based on Normal-Span Violations.}
The signs assigned in Step 1 may be incorrect because of numerical errors and incomplete recovery.
We check them against the collected decision-boundary normals: with accurate recovery and correct signs, each normal lies in the span of the signatures classified as active on its corresponding side, after they are pulled back to the input space.
We count violations of this condition and iteratively flip the sign that most reduces the count, stopping when no single sign flip improves it.
Numerical errors and unrecovered neurons may leave residual violations even with correct signs.
\Cref{app:sign-correction} gives the definitions and detailed correction procedure.

\subsection{Experiments}
\label{subsec:e2e-experiments}
\begin{table}[tb]
\caption{Neuron recovery, queries, and running times for MNIST- and FMNIST-trained models with four or six hidden layers of width 16 (Int.: intersection).}
\label{tab:e2e-extraction-results}
    \centering
\addtolength{\tabcolsep}{0.4pt}
\scalebox{0.8}{
\begin{tabular}{llrrrrrrrrr}
\toprule
Model & Metric & \shortstack{Int. space\\collection} & L1 & L2 & L3 & L4 & L5 & L6 & L7 & \shortstack{Entire\\model} \\
\midrule
\multirow{5}{*}{\shortstack{MNIST\\784-$16^{(4)}$-10}} & Active neurons & -- & 16 & 16 & 16 & 16 & 10 & -- & -- & 74 \\
 & Recovered neurons & -- & 16 & 16 & 16 & 16 & 10 & -- & -- & 74 \\
 & Signature time & -- & 2.95s & 4.20s & 9.02s & 5.15s & 3.20s & -- & -- & 25s \\
 & Sign time & -- & 0.23s & 44s & 29s & 31s & 0.23ms & -- & -- & 1m44s \\
 & Total queries & $2^{32.31}$ & $2^{17.10}$ & $2^{17.10}$ & $2^{18.42}$ & $2^{17.69}$ & $0$ & -- & -- & $2^{32.31}$ \\
\midrule
\multirow{5}{*}{\shortstack{MNIST\\784-$16^{(6)}$-10}} & Active neurons & -- & 16 & 16 & 15 & 15 & 15 & 16 & 10 & 103 \\
 & Recovered neurons & -- & 16 & 16 & 15 & 15 & 15 & 16 & 10 & 103 \\
 & Signature time & -- & 4.72s & 17s & 17m14s & 41m05s & 4m09s & 1m42s & 1m05s & 1h05m36s \\
 & Sign time & -- & 0.87s & 4.89s & 6m45s & 7m06s & 6m06s & 6m32s & 0.15ms & 26m34s \\
 & Total queries & $2^{36.62}$ & $2^{17.10}$ & $2^{17.10}$ & $2^{20.44}$ & $2^{21.90}$ & $2^{21.75}$ & $2^{21.82}$ & 0 & $2^{36.62}$ \\
\midrule
\multirow{5}{*}{\shortstack{FMNIST\\784-$16^{(4)}$-10}} & Active neurons & -- & 16 & 16 & 16 & 16 & 10 & -- & -- & 74 \\
 & Recovered neurons & -- & 16 & 16 & 16 & 16 & 10 & -- & -- & 74 \\
 & Signature time & -- & 2.99s & 5.39s & 59s & 1m21s & 3.23s & -- & -- & 2m32s \\
 & Sign time & -- & 0.24s & 35s & 39s & 33s & 0.14ms & -- & -- & 1m48s \\
 & Total queries & $2^{32.30}$ & $2^{17.10}$ & $2^{17.11}$ & $2^{21.71}$ & $2^{22.24}$ & 0 & -- & -- & $2^{32.30}$ \\
\midrule
\multirow{5}{*}{\shortstack{FMNIST\\784-$16^{(6)}$-10}} & Active neurons & -- & 16 & 16 & 16 & 15 & 16 & 16 & 10 & 105 \\
 & Recovered neurons & -- & 16 & 16 & 16 & 15 & 16 & 16 & 10 & 105 \\
 & Signature time & -- & 4.79s & 16s & 21s & 39m25s & 1m07s & 4m45s & 1m05s & 47m04s \\
 & Sign time & -- & 0.90s & 4.45s & 10m05s & 6m14s & 6m34s & 7m17s & 0.15ms & 30m16s \\
 & Total queries & $2^{36.62}$ & $2^{17.10}$ & $2^{17.09}$ & $2^{17.47}$ & $2^{21.80}$ & $2^{21.26}$ & $2^{23.67}$ & 0 & $2^{36.62}$ \\
\bottomrule
\end{tabular}
}
\end{table}

We evaluate end-to-end hard-label extraction on MNIST- and Fashion-MNIST (FMNIST)-trained models, reporting recovered neuron counts, query counts, running times, and output agreement with the oracle. The target models are fully connected ReLU networks with four or six hidden layers of 16 neurons each, 784 inputs, and 10 outputs.
We initialized the weights using Kaiming normal initialization and trained the models for 30 epochs using the Adam optimizer with a learning rate of $10^{-3}$.
We used cross-entropy loss for 10-class classification.

The attack collects \intersectionspaces before recovering layers from shallow to deep.
We collected 20,000 \intersectionspaces for each model with four hidden layers and 400,000 for each model with six hidden layers.
We report the collection query cost separately.
For sign recovery, we modify \cref{alg:signature-weighted-sign-recovery} to use $D=|z|-|z'|$ and divide each competitor $\tilde{\bmw}_j$ by its input-space pullback norm $\|\bfGamma^\top(\alpha_j\bmw_j)\|_2$ before forming the correction matrix, omitting zero pullbacks.
These changes slightly improved performance.

\Cref{tab:e2e-extraction-results} summarizes the results.
No persistent neurons were identified in these experiments, so we did not perform their recovery.
However, we observed dead neurons, which were always inactive, and almost-dead neurons, whose activation probability was at most $1\%$ under standard Gaussian inputs.
The Active neurons rows exclude these neurons; for the output layer, they count all 10 output units, to which ReLU is not applied.
Recovered neurons gives the number of neurons successfully recovered.
Signature time and Sign time report the running times for signature and sign recovery, respectively.
In the Total queries row, Int. space collection reports the cost of collecting \intersectionspaces, L1--L7 report the additional queries for each layer, and Entire model includes both costs.
Sign recovery reuses the collected \intersectionspaces and requires zero additional queries, so all layer-recovery queries are for signature recovery.

For all four models in the table, we recovered every neuron except dead and almost-dead neurons.
The maximum $L_2$ distance between an estimated hidden-layer signature and the corresponding true weight vector, after matching neurons, resolving the sign, and normalizing each vector, was $9.34 \times 10^{-6}$.
Initial \intersectionspace collection dominated the query cost; signature recovery required relatively few additional queries.

We measured output agreement as the fraction of 100,000 standard Gaussian inputs drawn from $\mathcal{N}(\bm 0,\bfI_{784})$ for which the extracted model and oracle predicted the same class.
The agreement rates were $98.477\%$ and $99.948\%$ for the MNIST models with four and six hidden layers, respectively, and $100.000\%$ and $99.744\%$ for the corresponding FMNIST models.
Numerical errors and unrecovered almost-dead neurons are possible sources of the remaining disagreements.

%% file: 90_appendix.tex
\section{Recent Results Using an Artificial Model \cite{DBLP:conf/eurocrypt/CarliniCHRS25}}
\label{appsec:artificial}

Very recently, the ePrint version of \cite{DBLP:conf/eurocrypt/CarliniCHRS25} was updated to include a black-box end-to-end hard-label model-extraction demonstration~\cite{DBLP:journals/iacr/CarliniCHRS24}.
The authors constructed a toy model with 32 inputs, three hidden layers of width 32, and 10 outputs.
They provide black-box procedures for collecting \intersectionpoints, grouping them, recovering signatures, and recovering signs.
By combining these procedures, they demonstrate the feasibility of end-to-end model extraction.

However, this result should be interpreted carefully.
Their target model is neither trained nor conventionally initialized.
Its hidden-layer weights are orthogonal, and its biases are adjusted so that each hidden neuron is active for about half of the sampled inputs.
Thus, it is a highly structured artificial model to demonstrate the feasibility of black-box hard-label model extraction.

This construction substantially reduces the difficulty of sign recovery.
Their black-box implementation first attempts to find \intersectionpoints at which all neurons in the layers preceding the target layer are active.
Once such an \intersectionpoint is found, the local linear map is full rank and isotropic in this model.
Thanks to orthogonal weights, the off-side control direction is orthogonal to the off-side decision-boundary normal. 
Therefore, when the signatures and decision-boundary normals are recovered with sufficient precision, walking along the off side never changes the values in subsequent layers before a past toggle occurs. 
Consequently, no future toggle occurs on the off side.
If a future toggle is detected before a past toggle, the corresponding direction can be identified as the on side.

Extending this technique beyond the artificial model is nontrivial.
Without such a special artificial model, \intersectionpoints at which all preceding neurons are active are generally difficult to find reliably, whether the model is trained or conventionally initialized.
The orthogonal weights, the activation-rate calibration, and the small number of hidden layers make such \intersectionpoints considerably easier to find.
Even in the updated ePrint version~\cite{DBLP:journals/iacr/CarliniCHRS24}, the experiments on a trained CIFAR-10 model remain white-box evaluations that use \(10^2\)--\(10^3\) \intersectionpoints.
Thus, the feasibility of a black-box implementation for general non-artificial models remains to be verified.

\section{Collecting \intersectionspaces}\label{appsec:collect_dual_spaces}

\begin{figure}[t]
    \centering
    \includegraphics[width=0.65\linewidth]{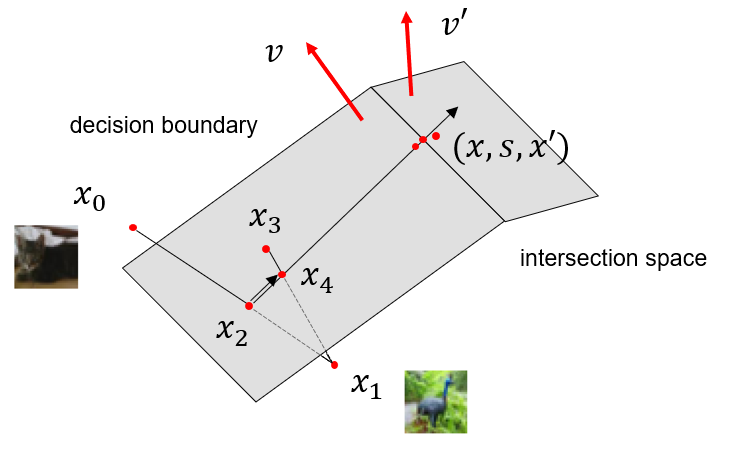}
    \caption{Overview of how to collect \intersectionpoints and \intersectionspaces.}
    \label{fig:overview-of-collecting}
\end{figure}

In hard-label model extraction, the attacker first collects a sufficient number of \intersectionspaces.
In this section, we describe how to collect the \intersectionspaces.
An outline of this procedure is shown in \cref{fig:overview-of-collecting}.

\paragraph{Step 1: Find a point $\bmx_2$ on the decision boundary.}
We sample $\bmx_0, \bmx_1 \in \R^{d_0}$ from an adequate distribution (e.g., a Gaussian distribution) such that $f(\bmx_0) \neq f(\bmx_1)$, and perform a binary search between them to obtain a boundary point $\bmx_2$.

\paragraph{Step 2: Find another point $\bmx_4$ on the decision boundary.}
We move from $\bmx_2$ in a random direction to obtain $\bmx_3$ sufficiently close to the same decision boundary.
Without loss of generality, assume $f(\bmx_3) = f(\bmx_0)$.
A second boundary point $\bmx_4$ is obtained by a binary search between $\bmx_1$ and $\bmx_3$.

\paragraph{Step 3: Move along the decision boundary until reaching an \intersectionpoint $\bms$.}
Let $\Delta \bmx := \bmx_4 - \bmx_2$ and
$\bmx_{\gamma} := \bmx_2 + \gamma \Delta \bmx$
for a small $\gamma$.
At this stage, we first recover the normal vector $\bmv$ of the local decision boundary
and use it to guide the movement so that $\bmx_{\gamma}$ follows the decision boundary
exactly along a straight line.
As $\bmx_{\gamma}$ moves along the decision boundary, it eventually crosses an
activation boundary of the ReLU network.
The crossing point, called an \intersectionpoint, is denoted by $\bms$.
After crossing, the point is projected back onto the decision boundary to obtain
$\bmx$, while the point immediately before crossing is denoted by $\bmx'$.
This yields a triplet $(\bmx,\bms,\bmx')$.

\paragraph{Step 4: Find the normal vectors $\bmv$ and $\bmv'$.}
For a boundary point $\bmx \in \R^{d_0}$ and a slight perturbation $\bmx + \alpha \bme_1$, using the standard basis vectors $\bme_1, \ldots, \bme_{d_0}$, we find $\beta_i$ such that $\bmx + \alpha \bme_1 + \beta_i \bme_i$ lies on the decision boundary.
The normal vector is aligned with $(1/\beta_1, \ldots, 1/\beta_{d_0})^{\top}$ and normalized.
Applying this procedure at $\bmx$ and $\bmx'$ gives the unit normals $\bmv$ and $\bmv'$.

\paragraph{Step 5: Define the \intersectionspace.}
The dual space is the intersection of the two local decision-boundary hyperplanes
determined by $\bmv$ and $\bmv'$, or equivalently, the orthogonal complement of
their span.
The \intersectionspace $\mathcal{S}$ associated with $\bms$ is defined as $\mathcal{S} :=\bms + \Span\{\bmv,\bmv'\}^{\perp} \subset \R^{d_0}$.
We represent $\mathcal{S}$ by the triplet $(\bms,\bmv,\bmv')$.
In particular, let
$\bmn_1,\ldots,\bmn_{d_0-2}$
be an orthonormal basis of
$\Span\{\bmv,\bmv'\}^{\perp}$.
We then construct the basis matrix
\[
    \bfN
    :=
    \begin{bmatrix}
        \bmn_1 & \cdots & \bmn_{d_0-2}
    \end{bmatrix}
    \in \R^{d_0 \times (d_0-2)}.
\]
Then, every point $\bmx \in \mathcal{S}$ can be parameterized as
$\bmx = \bms + \bfN \bmu,   \bmu \in \R^{d_0-2}$.

\paragraph{Step 6: Repeat these steps.}
Repeating the procedure yields \intersectionspaces
$\mathcal{S}_1, \ldots, \mathcal{S}_N$, which are used in \cite{DBLP:conf/eurocrypt/CarliniCHRS25} to reconstruct the weight parameters.

\section{Pseudocode for Signature Recovery}\label{appsec:signature-recovery}

The following algorithms formalize the consistency test and full signature recovery in \cref{subsec:recovering_the_signature}.
\textsc{IsConsistent} tests for a shared affine constraint, which need not correspond to a target-layer neuron.
\textsc{RecoverSignature} returns a unit signature and its consistently scaled bias only when the full affine constraint has a one-dimensional normal space.
In numerical implementations, ranks and null spaces require a tolerance; the formulas below describe exact arithmetic.

\begin{algorithm}[htb]
\caption{\textsc{IsConsistent}}
\label{alg:pre-Consistent}
\begin{algorithmic}[1]
\Require $\mathcal{S}_1,\mathcal{S}_2$; recovered weights and biases through layer $\ell-1$
\Ensure whether the pair passes the affine consistency test
\For{$i\in\{1,2\}$}
    \State Read $(\bms_i,\bfN_i)$ from $\mathcal{S}_i$.
    \State $\bfT_i\gets\bfGamma_{\bms_i}^{(\ell-1)}\bfN_i$; $\bmu_i\gets\bmh^{(\ell-1)}(\bms_i)$.
\EndFor
\State $\nu\gets$ number of layer-$(\ell-1)$ neurons active at either point.
\If{$\ell=1$}
    \State $\nu\gets d_0$.
\EndIf
\State \Return $\operatorname{rank}[\bfT_1,\bfT_2,\bmu_1-\bmu_2]<\nu$.
\end{algorithmic}
\end{algorithm}

\begin{algorithm}[htb]
\caption{\textsc{RecoverSignature}}
\label{alg:pre-signaturerec}
\begin{algorithmic}[1]
\Require $\mathcal{S}_1,\ldots,\mathcal{S}_m$ for a candidate neuron in layer $\ell$; recovered preceding-layer weights and biases
\Ensure a unit signature $\bmw$ and bias $b$, or $\bot$ if not uniquely determined
\For{$i=1,\ldots,m$}
    \State Read $(\bms_i,\bfN_i)$ from $\mathcal{S}_i$.
    \State $\bfT_i\gets\bfGamma_{\bms_i}^{(\ell-1)}\bfN_i$; $\bmu_i\gets\bmh^{(\ell-1)}(\bms_i)$.
\EndFor
\State $\bfU\gets[\bfT_1,\ldots,\bfT_m,\bmu_2-\bmu_1,\ldots,\bmu_m-\bmu_1]$.
\If{$\operatorname{rank}(\bfU)\ne d_{\ell-1}-1$}
    \State \Return $\bot$. \Comment{Insufficient constraints or inconsistent data}
\EndIf
\State Choose a unit vector $\bmw\in\ker(\bfU^\top)$.
\State $b\gets-\bmw^\top\bmu_1$.
\State \Return $(\bmw,b)$.
\end{algorithmic}
\end{algorithm}

%% file: 91_on_signature.tex
\clearpage
\section{On-Side Variance in Signature-Weighted Cosine}
\label{sec:on-side-amp}

Let $\bfA$ be 
\begin{align}
\bfA=[ \tilde \bmw_1,\ldots, \tilde \bmw_{t-1}, \tilde \bmw_{t+1},\ldots, \tilde \bmw_{d_\ell}],
\end{align}
Then, 
\begin{align}
&\tilde \bfC:=\sum_{j\in \mathcal{J}_{\mathrm{comp}}} (\alpha_j \tilde \bmw_j)  (\alpha_j \tilde \bmw_j)^\top 
=\sum_{j\in \mathcal{J}_{\mathrm{comp}}} \tilde \bmw_j \tilde \bmw_j^\top = \bfA \bfA^\top.
\end{align}
We assume $\operatorname{rank}(\tilde\bfC)=r_{\bfGamma}$ and interpret $\tilde\bfC^{-1}$ on $\operatorname{Im}(P_{\bfGamma})$.
Then, we use the approximation
\[
z_{\mathrm{on}}
\approx z_{\mathrm{off}}+
\frac{\alpha_t \beta_t\|\tilde \bfC^{-1/2}\tilde \bmw_t\|}
{\|\tilde \bfC^{-1/2}\tilde \bmv_{\mathrm{off}}\|}.
\]

We first consider $\|\tilde \bfC^{-1/2}\tilde \bmv_{\mathrm{off}}\|^2 =\tilde \bmv_{\mathrm{off}}^\top\tilde \bfC^{-1}\tilde \bmv_{\mathrm{off}}$.
Let $\bma\in\R^{|\mathcal{J}_{\mathrm{comp}}|}$ be a zero-padded coefficient vector such that $\beta_j = a_j$ for $j \in \mathcal{J}_{\mathrm{act}}$ and $a_j=0$ otherwise. Since $\tilde \bmv_{\mathrm{off}}=\bfA\bma$,
\[
\|\tilde \bfC^{-1/2}\tilde \bmv_{\mathrm{off}}\|^2
=\bma^\top\left(\bfA^\top(\bfA \bfA^\top)^{-1}\bfA \right)\bma
=\bma^\top \bfP_{\bfA} \bma.
\]
Here, $\bfP_\bfA$ is an orthogonal projection matrix of rank $r_{\bfGamma}$ in the $|\mathcal{J}_{\mathrm{comp}}|$-dimensional coefficient space. Therefore, if the direction of $\bma$ is not biased relative to the row space of $\bfA$,
\[
\|\tilde \bfC^{-1/2}\tilde \bmv_{\mathrm{off}}\|^2
\approx\frac{r_{\bfGamma}}{|\mathcal{J}_{\mathrm{comp}}|}
\sum_{j\in \mathcal{J}_{\mathrm{act}}}\beta_j^2.
\]

Next, we consider $\|\tilde \bfC^{-1/2}\tilde \bmw_t\|^2=\tilde \bmw_t^\top\tilde \bfC^{-1}\tilde \bmw_t$. If $\tilde \bmw_t$ were included among the generators of the competitor space, it would be expected to have the same value $r_{\bfGamma}/|\mathcal{J}_{\mathrm{comp}}|$ as the denominator.
However, the actual $\tilde \bfC$ does not contain the target vector $\tilde \bmw_t$. 
If $\tilde \bmw_t$ has comparable components in each eigendirection of the competitor space, sparse directions are amplified and dense directions are weakened. 
For comparison, consider a reference model in which the target and competitors are independent Gaussian samples with the same covariance. By a property of the inverse of a Wishart matrix,
\[
\E\left[\tilde \bmw_t^\top\tilde \bfC^{-1}\tilde \bmw_t\right]
=\frac{r_{\bfGamma}}{|\mathcal{J}_{\mathrm{comp}}|-r_{\bfGamma}-1}.
\]
In practice, however, the target of model extraction consists only of fixed parameters, and trained models may have biases in their weights. 
Therefore, this theoretical assumption does not generally hold. 
Instead, we measure the quantity directly at each \intersectionspace and compare it. 
We used the same random and MNIST-trained models. 
\begin{table}[t]
\centering
\caption{Measured values of $\tilde \bmw_t^\top\tilde \bfC^{-1}\tilde \bmw_t$ and the reference-model prediction.}
\label{tab:signature-weighted-target-metric}
\begin{tabular}{c|cc|cc|c} \toprule
$r_{\bfGamma}$ & \multicolumn{2}{c|}{Random} & \multicolumn{2}{c|}{MNIST} & Prediction \\
& 2nd layer & 4th layer & 2nd layer & 4th layer & \\ \midrule
50 & 0.6786 & 0.6602 & 0.6714 & 0.6699 & 0.6579 \\
55 & 0.7629 & 0.7648 & 0.7500 & 0.7713 & 0.7746 \\
60 & 0.9057 & 0.9049 & 0.9027 & 0.9012 & 0.9091 \\
65 & 1.0550 & 1.0758 & 1.0501 & 1.0879 & 1.0656 \\
70 & 1.2378 & 1.3441 & 1.2487 & -- & 1.2500 \\
75 & 1.4165 & -- & 1.4326 & -- & 1.4706 \\ \bottomrule
\end{tabular}
\end{table}
\Cref{tab:signature-weighted-target-metric} summarizes $\tilde \bmw_t^\top\tilde \bfC^{-1}\tilde \bmw_t$ for varying values of $r_{\bfGamma}$. 
As a result, the estimate obtained from the above reference model remains a good approximation.

The typical squared ratio is estimated as
\[
\left( \frac{\alpha_t \beta_t\|\tilde \bfC^{-1/2}\tilde \bmw_t\|}
{\|\tilde \bfC^{-1/2}\tilde \bmv_{\mathrm{off}}\|}
\right)^2
\approx
\frac{|\mathcal{J}_{\mathrm{comp}}|}{|\mathcal{J}_{\mathrm{comp}}|-r_{\bfGamma}-1}
\frac{\beta_t^2}{\sum_{j\in \mathcal{J}_{\mathrm{act}}}\beta_j^2}.
\]
For the $\bfGamma$-weighted cosine method, the on-side variance was estimated to exceed the off-side variance by $\frac{\beta_t^2}{\sum_{j\in \mathcal{J}_{\mathrm{act}}}\beta_j^2}\approx\frac{1}{|\mathcal{J}_{\mathrm{act}}|}$. For the signature-weighted cosine method, this amplification is multiplied by
\[
\frac{|\mathcal{J}_{\mathrm{comp}}|}{|\mathcal{J}_{\mathrm{comp}}|-r_{\bfGamma}-1}\approx 2.
\]
Consequently, under a normal approximation,
\begin{align}
z_{\mathrm{on}}
\overset{\mathrm{approx}}{\sim}\mathcal{N}\!\left(0,\frac{1}{r_{\bfGamma}}+\frac{4}{d_\ell}\right).\notag
\end{align}



%% file: 92_e2e_appendix.tex
\section{Derivation of Noise Components under Varying Activation Patterns}
\label{app:noise-component-derivation}

This section derives the mechanism described in case (ii) of \cref{sec:e2e}: \intersectionspaces can pass the consistency check even when the $j$th neuron in layer $\ell$ has different activation states at their representative points.
Suppose that each \intersectionspace corresponds to the $k$th neuron in layer $\ell+1$, and that the state of the $j$th neuron in layer $\ell$ is fixed in a neighborhood of each representative point.
The class pair defining the decision boundary and the activation patterns from layer $\ell$ onward are shared, except for the $j$th neuron in layer $\ell$ and the $k$th neuron in layer $\ell+1$.
Activation patterns in layers preceding $\ell$ may vary freely.
Under these conditions, \intersectionspaces whose representative points have different activation states for the $j$th neuron in layer $\ell$ can pass the consistency check together.

Let $\mathcal{J}$ be the index set of neurons in layer $\ell$, excluding the $j$th neuron, that are active at all the points under consideration.
The pre-activation of the $k$th neuron in the next layer is
\begin{align}
    \hat h_k^{(\ell+1)}
    &=\sum_{r\in\mathcal{J}}w_{k,r}^{(\ell+1)}h_r^{(\ell)}
    +b_k^{(\ell+1)}+w_{k,j}^{(\ell+1)}h_j^{(\ell)}
\end{align}
To describe the decision boundary, we work with logits before softmax.
Following \cref{subsec:relu_network}, consider an $L$-layer ReLU network whose final-layer output $\bmh^{(L)}=\bfW^{(L)}\bmh^{(L-1)}+\bmb^{(L)}$ is the logit vector.
Thus, the final layer considered here is a linear layer with a bias, without softmax.
Applying softmax to the logits yields class probabilities, but preserves their ordering, including ties.
Hence, the decision boundary of a classifier that selects the class with the largest logit can be described using logit differences, without applying softmax.

We derive the decision-boundary equation using the logit difference on the side where the $k$th neuron in layer $\ell+1$ is inactive.
We are concerned with intersections on this neuron's activation boundary, where $\hat h_k^{(\ell+1)}=0$ and its output is zero.
The expressions from the active and inactive sides therefore agree on the intersection, so the inactive-side expression suffices.
Let $\bfD^{(\ell+1)},\ldots,\bfD^{(L-1)}$ be the downstream activation patterns on this side; in particular, the $k$th diagonal entry of $\bfD^{(\ell+1)}$ is zero.
With these patterns fixed, the composition of the downstream layers is
\begin{align}
    \bmh^{(L)}
    &=\mathbf B^{(\ell)}\bmh^{(\ell)}+\bm c^{(\ell)},\\
    \mathbf B^{(\ell)}
    &=\bfW^{(L)}\bfD^{(L-1)}\bfW^{(L-1)}
    \cdots\bfD^{(\ell+1)}\bfW^{(\ell+1)}
\end{align}
Here, $\bm c^{(\ell)}$ is the constant vector obtained by composing the biases from layer $\ell+1$ onward.
Specifically, initialize $\mathbf B^{(L-1)}=\bfW^{(L)}$ and $\bm c^{(L-1)}=\bmb^{(L)}$, and compute, for $r=L-1,\ldots,\ell+1$,
\begin{align}
    \mathbf B^{(r-1)}&=\mathbf B^{(r)}\bfD^{(r)}\bfW^{(r)},\\
    \bm c^{(r-1)}&=\mathbf B^{(r)}\bfD^{(r)}\bmb^{(r)}+\bm c^{(r)}
\end{align}
Let $c,c'$ be the classes defining the decision boundary.
Their logit difference is
\begin{align}
    h_c^{(L)}-h_{c'}^{(L)}
    &=(\bme_c-\bme_{c'})^\top\mathbf B^{(\ell)}\bmh^{(\ell)}
    +(\bme_c-\bme_{c'})^\top\bm c^{(\ell)}
\end{align}
where $\bme_c$ and $\bme_{c'}$ are the corresponding standard basis vectors.
Define
\begin{align}
    \bm\beta&=\mathbf B^{(\ell)\top}(\bme_c-\bme_{c'}), &
    b_{\mathrm{down}}&=(\bme_c-\bme_{c'})^\top\bm c^{(\ell)}
\end{align}
Then $\beta_r$ is the coefficient describing the contribution of $h_r^{(\ell)}$ to the logit difference, and $b_{\mathrm{down}}$ accounts for the downstream biases.
The decision-boundary equation is therefore
\begin{align}
    \sum_{r\in\mathcal{J}}\beta_r h_r^{(\ell)}
    +b_{\mathrm{down}}+\beta_j h_j^{(\ell)}=0
\end{align}
In particular, $\beta_j$ is the coefficient of the output $h_j^{(\ell)}$ of the $j$th neuron in layer $\ell$ and is shared by both activation states.

Substitute $h_r^{(\ell)}=\bmw_r^{(\ell)\top}\bmh^{(\ell-1)}+b_r^{(\ell)}$ for each $r\in\mathcal{J}$, and collect the contributions of all neurons other than the $j$th neuron by defining
\begin{align}
    \tilde{\bmw}_k^{(\ell+1)}
    &=\sum_{r\in\mathcal{J}}w_{k,r}^{(\ell+1)}\bmw_r^{(\ell)}, &
    \tilde b_k^{(\ell+1)}
    &=\sum_{r\in\mathcal{J}}w_{k,r}^{(\ell+1)}b_r^{(\ell)}+b_k^{(\ell+1)},\\
    \bmv^{(\ell)}
    &=\sum_{r\in\mathcal{J}}\beta_r\bmw_r^{(\ell)}, &
    b_{\mathrm{DB}}^{(\ell)}
    &=\sum_{r\in\mathcal{J}}\beta_r b_r^{(\ell)}+b_{\mathrm{down}}
\end{align}
When the $j$th neuron in layer $\ell$ is inactive, $h_j^{(\ell)}=0$, so the intersection satisfies
\begin{align}
    \tilde{\bmw}_k^{(\ell+1)\top}\bmh^{(\ell-1)}+\tilde b_k^{(\ell+1)}&=0,\\
    \bmv^{(\ell)\top}\bmh^{(\ell-1)}+b_{\mathrm{DB}}^{(\ell)}&=0
\end{align}
When this neuron is active, $h_j^{(\ell)}=\hat h_j^{(\ell)}=\bmw_j^{(\ell)\top}\bmh^{(\ell-1)}+b_j^{(\ell)}$.
Substituting this expression yields
\begin{align}
    \tilde{\bmw}_k^{(\ell+1)\top}\bmh^{(\ell-1)}+\tilde b_k^{(\ell+1)}
    +w_{k,j}^{(\ell+1)}\hat h_j^{(\ell)}&=0,\\
    \bmv^{(\ell)\top}\bmh^{(\ell-1)}+b_{\mathrm{DB}}^{(\ell)}
    +\beta_j\hat h_j^{(\ell)}&=0
\end{align}
Thus, in the input space of layer $\ell$, the activation-boundary normal and the decision-boundary normal acquire the additional terms $w_{k,j}^{(\ell+1)}\bmw_j^{(\ell)}$ and $\beta_j\bmw_j^{(\ell)}$, respectively.
However, multiplying the activation-boundary equation by $\beta_j$ and subtracting $w_{k,j}^{(\ell+1)}$ times the decision-boundary equation eliminates this neuron's contribution.
For both activation states, we obtain
\begin{align}
    \bigl(\beta_j\tilde{\bmw}_k^{(\ell+1)}
    -w_{k,j}^{(\ell+1)}\bmv^{(\ell)}\bigr)^\top\bmh^{(\ell-1)}
    +\beta_j\tilde b_k^{(\ell+1)}
    -w_{k,j}^{(\ell+1)}b_{\mathrm{DB}}^{(\ell)}=0
\end{align}
If this normal is nonzero, the \intersectionspaces for both activation states lie in a common hyperplane in the input space of layer $\ell$ and can therefore pass the consistency check.
Thus, even when both boundaries bend as the $j$th neuron in layer $\ell$ switches state, a shared linear constraint on their intersections remains.
Even if activation patterns in preceding layers differ, the tangent directions of each \intersectionspace, mapped to the input space of layer $\ell$ by the corresponding $\bfGamma_{\bms_i}^{(\ell-1)}$, are orthogonal to this common normal.
The activation patterns in preceding layers therefore need not be fixed.

\section{Cross-Layer Extraction of Unreachable Weights}
\label{app:cross-layer-unreachable}

We describe how cross-layer extraction~\cite{DBLP:conf/crypto/ItoMT26} applies to unreachable weights in the setting of \cref{sec:e2e}.
If a neuron with an unreachable weight is active at \intersectionpoints of a neuron in the next layer, the corresponding \intersectionspaces can be used for cross-layer recovery.
Let $k$ be the index of the target neuron in layer $\ell$, and suppose that its signature has been recovered except for coordinate $j$.
Consider the $\eta$th neuron in layer $\ell+1$, whose activation boundary satisfies $\bmw_\eta^{(\ell+1)\top}\bmh^{(\ell)}+b_\eta^{(\ell+1)}=0$.
Let $\mathcal{P}=[d_\ell]\setminus\{k\}$ be the index set of the other neurons, which have no unreachable weights in the setting considered here.
Let $\bmw_{\eta,\mathcal{P}}^{(\ell+1)}$ and $\bmh_{\mathcal{P}}^{(\ell)}$ be the restrictions of the next-layer weight vector and the layer-$\ell$ activation vector to $\mathcal{P}$, respectively.
The boundary equation becomes
\begin{align}
    \bmw_{\eta,\mathcal{P}}^{(\ell+1)\top}\bmh_{\mathcal{P}}^{(\ell)}
    +w_{\eta,k}^{(\ell+1)}h_k^{(\ell)}+b_\eta^{(\ell+1)}=0.
\end{align}
When the target neuron is active, $h_k^{(\ell)}=\bmw_k^{(\ell)\top}\bmh^{(\ell-1)}+b_k^{(\ell)}$.
Substituting this expression gives
\begin{align}
    &\bmw_{\eta,\mathcal{P}}^{(\ell+1)\top}\bmh_{\mathcal{P}}^{(\ell)}
    +w_{\eta,k}^{(\ell+1)}\bmw_k^{(\ell)\top}\bmh^{(\ell-1)}
    +w_{\eta,k}^{(\ell+1)}b_k^{(\ell)}+b_\eta^{(\ell+1)}=0.
\end{align}
Equivalently,
\begin{align}
    &\bigl(\bmw_{\eta,\mathcal{P}}^{(\ell+1)\top},
    w_{\eta,k}^{(\ell+1)}\bmw_k^{(\ell)\top}\bigr)
    \begin{pmatrix}\bmh_{\mathcal{P}}^{(\ell)}\\\bmh^{(\ell-1)}\end{pmatrix}
    +w_{\eta,k}^{(\ell+1)}b_k^{(\ell)}+b_\eta^{(\ell+1)}=0.
\end{align}
Thus, in the augmented space formed by concatenating the inputs to layers $\ell+1$ and $\ell$ as above, this equation defines an effective activation boundary.
Its normal is $(\bmw_{\eta,\mathcal{P}}^{(\ell+1)\top},w_{\eta,k}^{(\ell+1)}\bmw_k^{(\ell)\top})^\top$, and its bias is $w_{\eta,k}^{(\ell+1)}b_k^{(\ell)}+b_\eta^{(\ell+1)}$.
Applying the signature-recovery procedure for hard-label extraction to this boundary recovers the target signature multiplied by the next-layer weight.
Unlike the original cross-layer setting, all coordinates of $\bmw_k^{(\ell)}$ except coordinate $j$ have already been recovered.
These known coordinates allow us to determine $w_{\eta,k}^{(\ell+1)}$ and then recover the missing coordinate $w_{k,j}^{(\ell)}$.
The same approach extends to unreachable weights in multiple neurons within a layer.

\section{Intersection-Point Search Procedures}
\label{app:intersection-search-procedures}

\subsection{Searching for Informative Intersection Points}
\label{app:informative-intersection-search}

\paragraph{Problem setting.}
We describe the search for informative intersection points used to recover missing signature coordinates in \cref{sec:e2e}.
Consider the $k$th neuron in layer $\ell$.
Let $\{(\bms_i,\bmv_i,\bmv_i')\}_{i\in[m]}$ be the representations of the collected \intersectionspaces for this neuron, where $\bms_i$ is a representative \intersectionpoint and $\bmv_i,\bmv_i'$ are the adjacent decision-boundary normals.
Suppose that the $j$th neuron in layer $\ell-1$ is inactive at every collected point.
Thus, the $j$th input coordinate to layer $\ell$ is zero at $\bms_1,\ldots,\bms_m$.
Let $\Lambda=[d_{\ell-1}]\setminus\{j\}$ be the index set excluding the missing coordinate, and let $\bmw_{k,\Lambda}^{(\ell)}$ and $\bmh_{\Lambda}^{(\ell-1)}$ be the restrictions of the target signature and its input vector to $\Lambda$, respectively.
The target neuron's pre-activation is
\begin{align}
    \hat{h}_k^{(\ell)}
    &= \bmw_{k,\Lambda}^{(\ell)\top}\bmh_{\Lambda}^{(\ell-1)}
    + b_k^{(\ell)} + w_{k,j}^{(\ell)}h_j^{(\ell-1)}.
\end{align}
Here, $w_{k,j}^{(\ell)}$ is the $j$th coordinate of the target weight vector, and $h_j^{(\ell-1)}$ is the corresponding input coordinate.
At each collected \intersectionpoint, $h_j^{(\ell-1)}=0$ and the target neuron lies on its activation boundary, so
\begin{align}
    \bmw_{k,\Lambda}^{(\ell)\top}\bmh_{\Lambda}^{(\ell-1)}+b_k^{(\ell)}=0.
\end{align}
Let $\hat{h}_{\mathrm{part}}^{(\ell)}$ be the partial pre-activation obtained by omitting coordinate $j$:
\begin{align}
    \hat{h}_{\mathrm{part}}^{(\ell)}(\bmx)
    = \bmw_{k,\Lambda}^{(\ell)\top}\bmh_{\Lambda}^{(\ell-1)}(\bmx)+b_k^{(\ell)}.
\end{align}
Before sign recovery, the orientation of the recovered activation boundary remains ambiguous.\footnote{The signs of $\bmw_k^{(\ell)}$ and $b_k^{(\ell)}$ have not yet been recovered. This sign ambiguity does not affect the search along the activation boundary described below, which does not require distinguishing its active and inactive sides.}

\paragraph{Searching for informative intersection points.}
The first approach searches for an \intersectionpoint at which $h_j^{(\ell-1)}>0$.
Choose an existing \intersectionpoint of the target neuron, say $\bms_1$, and search along the intersection for a point $\bmx_0$ satisfying $\hat{h}_j^{(\ell-1)}(\bmx_0)=0$.
By definition, $\bms_1$ lies on the model's decision boundary and satisfies $\hat{h}_k^{(\ell)}(\bms_1)=0$.
Since the weights and signs in preceding layers have already been recovered, we can evaluate $\hat{h}_j^{(\ell-1)}$.
The preceding neuron is inactive at the starting point, with $\hat{h}_j^{(\ell-1)}(\bms_1)<0$.

We move tangentially to both the activation boundary and the decision boundary, maintaining $\hat{h}_{\mathrm{part}}^{(\ell)}(\bmx)=0$ while remaining on the decision boundary, until $\hat{h}_j^{(\ell-1)}$ reaches zero.
To choose a direction, we project the ascent direction $\nabla_{\bmx}\hat{h}_j^{(\ell-1)}(\bmx)$ onto the orthogonal complement of the span of two normals: the activation-boundary normal $\nabla_{\bmx}\hat{h}_{\mathrm{part}}^{(\ell)}(\bmx)$ and either adjacent decision-boundary normal, $\bmv_1$ or $\bmv_1'$.
If the search crosses a recovered activation boundary in a preceding layer, the local linear map changes, and we update the direction accordingly.
Crossing an activation boundary in a subsequent layer can also invalidate the direction, but detecting such a crossing is difficult because that layer has not yet been recovered.
If the search fails because of such a crossing, we restart from another \intersectionpoint of the target neuron.

Once a point $\bmx_0$ satisfying $\hat{h}_j^{(\ell-1)}(\bmx_0)=0$ is found on the target intersection, we perturb it to obtain a point $\bmx_1$ with $\hat{h}_j^{(\ell-1)}(\bmx_1)=\epsilon$, where $\epsilon$ is a sufficiently small positive value.
The perturbed point remains close to both boundaries, while the input coordinate corresponding to the missing weight is now nonzero.
Starting from $\bmx_1$, we search for an \intersectionpoint using the procedure described in \cref{appsec:collect_dual_spaces}.
This may yield a point that reveals the missing coordinate.
If the attempt fails, we restart from another \intersectionpoint associated with the target neuron.
In our experiments, we try at most 16 starting \intersectionpoints.

\subsection{Searching for Intersections to Validate Candidate Signatures}
\label{app:candidate-intersection-search}

We describe the intersection search used to validate candidate signatures in \cref{sec:e2e}.
Suppose that the signatures and signs in all layers preceding layer $\ell$ have been recovered.
Let $\bmw$ and $b$ be the candidate signature and its bias, respectively.
The candidate pre-activation is $\hat{h}(\bmx)=\bmw^\top\bmh^{(\ell-1)}(\bmx)+b$.

We reuse the \intersectionpoints already collected for signature recovery.
Let $\bms_1,\ldots,\bms_m$ be these points after excluding those used to recover the candidate signature, since reusing the latter would not provide an independent check.
Each \intersectionpoint is adjacent to two local pieces of the decision boundary.
For each $\bms_i$, let $\bmx_i$ and $\bmx_i'$ be points on these two pieces, and let $\bmv_i$ and $\bmv_i'$ be their respective normals.
Choose one of the pieces, say the one containing $\bmx_i$ with normal $\bmv_i$.
Let $\bfGamma_i^{(\ell-1)}$ be the forward local linear matrix through layer $\ell-1$ at $\bmx_i$.
The input-space normal of the candidate activation boundary is $\nabla_{\bmx}\hat{h}(\bmx_i)=\bfGamma_i^{(\ell-1)\top}\bmw$.
We seek the intersection by moving toward the candidate activation boundary while remaining on the chosen decision-boundary piece.

The pre-activation gradient projected onto the tangent space of the decision boundary is
\begin{align}
    \nabla_{\parallel\mathrm{DB}}\hat{h}(\bmx_i)
    =\bfGamma_i^{(\ell-1)\top}\bmw
    -\frac{\bmv_i\bigl(\bmv_i^\top\bfGamma_i^{(\ell-1)\top}\bmw\bigr)}{\|\bmv_i\|_2^2}.
\end{align}
When this projected gradient is nonzero, moving against it if $\hat{h}(\bmx_i)>0$, or along it if $\hat{h}(\bmx_i)<0$, brings us closer to the candidate boundary.
If the intersection can be reached without crossing another activation boundary, the required displacement is
\begin{align}
    \Delta\bmx_i
    =-\frac{\hat{h}(\bmx_i)}{\|\nabla_{\parallel\mathrm{DB}}\hat{h}(\bmx_i)\|_2}
    \frac{\nabla_{\parallel\mathrm{DB}}\hat{h}(\bmx_i)}{\|\nabla_{\parallel\mathrm{DB}}\hat{h}(\bmx_i)\|_2}.
\end{align}
The corresponding distance is $|\hat{h}(\bmx_i)|/\|\nabla_{\parallel\mathrm{DB}}\hat{h}(\bmx_i)\|_2$.
Under this condition, $\bmx_i+\Delta\bmx_i$ lies at the intersection of the decision boundary and the candidate activation boundary.
We then check for a bend in the decision boundary to assess the candidate signature.

In practice, we compute the predicted distance $|\hat{h}(\bmx_i)|/\|\nabla_{\parallel\mathrm{DB}}\hat{h}(\bmx_i)\|_2$ for each starting point $\bmx_i$ and attempt the searches in ascending order of this distance.
Prioritizing nearby intersections reduces the likelihood of crossing other ReLU cells.
Empirically, this procedure reaches the desired intersection in most cases.

\section{Sign Correction Using Decision-Boundary Normals}
\label{app:sign-correction}

We detail Step 2 of the sign-recovery procedure in \cref{sec:e2e}.
The signs assigned in Step 1 are not necessarily correct because numerical errors and incomplete recovery can affect the underlying sign estimates.
In this stage, we check their consistency with the decision-boundary normals already collected and revise the assignments to improve that consistency.
The key property is that each decision-boundary normal lies in the span of the signatures of neurons active on its corresponding side, after these signatures are pulled back to the input space.

Let $m$ be the number of \intersectionspaces collected for signature recovery in the same layer.
The representation of each space includes one normal from each of the two decision-boundary pieces adjacent to its representative point, giving $2m$ normals in total.
Let $\bmv_1,\ldots,\bmv_{2m}$ be these normals.
For each normal, we use the activation states at the nearby point on the side from which it was obtained.

Let $\mathcal{J}_{\mathrm{rec}}$ be the index set of recovered neurons in the layer, and let $\bm\alpha$ be their current sign assignment.
Let $\mathcal{J}_{\mathrm{inactive},i}(\bm\alpha)$ be the subset classified as inactive under $\bm\alpha$ at the nearby point where $\bmv_i$ was obtained.
Following the candidate-set construction in \cref{sec:sign}, define $\mathcal{J}_{\mathrm{cand},i}(\bm\alpha)=\mathcal{J}_{\mathrm{rec}}\setminus\mathcal{J}_{\mathrm{inactive},i}(\bm\alpha)$.
Here every recovered neuron has an assigned sign, so this candidate set consists of neurons classified as active.
Unlike $\mathcal{J}_{\mathrm{comp}}$ in single-neuron sign recovery, $\mathcal{J}_{\mathrm{rec}}$ excludes no target neuron: each normal is checked using all recovered neurons classified as active on its side.
Let $\bfGamma_i$ be the forward local linear matrix to the target layer's input at that point.
The signature $\bmw_j$ pulls back to the input-space normal $\bfGamma_i^\top\bmw_j$.
Since the observed normal $\bmv_i$ is also an input-space vector, we compare it with the span $\mathcal{V}_i(\bm\alpha) =\operatorname{span}\{\bfGamma_i^\top\bmw_j \mid j\in\mathcal{J}_{\mathrm{cand},i}(\bm\alpha)\}$.
Flipping a sign changes not only the orientation of a signature but also which neurons are classified as active and hence included in this span.
If all contributing neurons have been recovered accurately and their signs are correct, each normal lies in its corresponding span.

We define the \emph{violation count} as the number of normals that lie outside their corresponding spans: $V(\bm\alpha) =\#\{i\in[2m] \mid \bmv_i\notin\mathcal{V}_i(\bm\alpha)\}$.
We use this count to update the signs as follows.
\begin{enumerate}
    \item \textbf{Evaluate the current assignment.}
    Initialize the signs using the result of Step 1.
    Construct $\mathcal{J}_{\mathrm{cand},i}(\bm\alpha)$ and its corresponding span for each normal, and compute $V(\bm\alpha)$.

    \item \textbf{Evaluate individual sign flips.}
    For each recovered neuron, consider the assignment obtained by flipping only its sign.
    Update $\mathcal{J}_{\mathrm{inactive},i}(\bm\alpha)$ and $\mathcal{J}_{\mathrm{cand},i}(\bm\alpha)$ and recompute the violation count for each candidate assignment.

    \item \textbf{Update or terminate.}
    If any candidate has a lower violation count than the current assignment, flip the sign that yields the largest reduction.
    Repeat the evaluation using the updated assignment.
    Terminate when no single sign flip reduces the violation count.
\end{enumerate}

Contributions from unrecovered almost-dead neurons and numerical errors may prevent the violation count from reaching zero even when the signs are correct.
Nevertheless, an appropriate sign assignment is expected to restore consistency with many of the observed normals and yield a small violation count.